\documentclass[ALICE,manyauthors]{cernphprep}
\usepackage[comma,square,numbers,sort&compress]{natbib}
\usepackage{hyperref}
\usepackage{lineno}
\usepackage{xspace}
\usepackage{xcolor}
\usepackage{rotating}
\usepackage{arydshln}
\usepackage[nolist]{acronym}
\usepackage{bm}
\usepackage[T1]{fontenc}
\usepackage{orcidlink}

\begin{document}
%

\newcommand{\pp}           {pp\xspace}
\newcommand{\ppbar}        {\mbox{$\mathrm {p\overline{p}}$}\xspace}
\newcommand{\XeXe}         {\mbox{Xe--Xe}\xspace}
\newcommand{\PbPb}         {\mbox{Pb--Pb}\xspace}
\newcommand{\pA}           {\mbox{p--A}\xspace}
\newcommand{\pPb}          {\mbox{p--Pb}\xspace}
\newcommand{\AuAu}         {\mbox{Au--Au}\xspace}
\newcommand{\dAu}          {\mbox{d--Au}\xspace}

\newcommand{\s}            {\ensuremath{\sqrt{s}}\xspace}
\newcommand{\snn}          {\ensuremath{\sqrt{s_{\text{\tiny NN}}}}\xspace}
\newcommand{\as}          {\ensuremath{\alpha_{\mathrm{S}}}\xspace}
\newcommand{\asQ}          {\ensuremath{\alpha_{\mathrm{S}}(Q^2)}\xspace}
\newcommand{\pt}           {\ensuremath{p_{\rm T}}\xspace}
\newcommand{\meanpt}       {$\langle p_{\mathrm{T}}\rangle$\xspace}
\newcommand{\ycms}         {\ensuremath{y_{\rm CMS}}\xspace}
\newcommand{\ylab}         {\ensuremath{y_{\rm lab}}\xspace}
\newcommand{\etarange}[1]  {\mbox{$\left | \eta \right |~<~#1$}}
\newcommand{\yrange}[1]    {\mbox{$\left | y \right |~<~#1$}}
\newcommand{\dndy}         {\ensuremath{\mathrm{d}N_\mathrm{ch}/\mathrm{d}y}\xspace}
\newcommand{\dndeta}       {\ensuremath{\mathrm{d}N_\mathrm{ch}/\mathrm{d}\eta}\xspace}
\newcommand{\avdndeta}     {\ensuremath{\langle\dndeta\rangle}\xspace}
\newcommand{\dNdy}         {\ensuremath{\mathrm{d}N_\mathrm{ch}/\mathrm{d}y}\xspace}
\newcommand{\Npart}        {\ensuremath{N_\mathrm{part}}\xspace}
\newcommand{\Ncoll}        {\ensuremath{N_\mathrm{coll}}\xspace}
\newcommand{\dEdx}         {\ensuremath{\textrm{d}E/\textrm{d}x}\xspace}
\newcommand{\RpPb}         {\ensuremath{R_{\rm pPb}}\xspace}
\newcommand{\XSec}         {cross-section\xspace}
\newcommand{\XSecs}        {cross-sections\xspace}
\newcommand{\RpA}          {\ensuremath{R_{\rm pA}}\xspace}
\newcommand{\RdAu}          {\ensuremath{R_{\rm dAu}}\xspace}
\newcommand{\RAA}          {\ensuremath{R_{\rm AA}}\xspace}
\newcommand{\RPbPb}        {\ensuremath{R_{\rm PbPb}}\xspace}
\newcommand{\otp}          {\ensuremath{\omega/\pi^0} ratio\xspace}

\newcommand{\nineH}        {$\sqrt{s}=0.9$~Te\kern-.1emV\xspace}
\newcommand{\seven}        {$\sqrt{s}=7$~Te\kern-.1emV\xspace}
\newcommand{\eight}        {$\sqrt{s}=8$~Te\kern-.1emV\xspace}
\newcommand{\thirteen}     {$\sqrt{s}=13$~Te\kern-.1emV\xspace}
\newcommand{\twoH}         {$\sqrt{s}=0.2$~Te\kern-.1emV\xspace}
\newcommand{\twosevensix}  {$\sqrt{s}=2.76$~Te\kern-.1emV\xspace}
\newcommand{\five}         {$\sqrt{s}=5.02$~Te\kern-.1emV\xspace}
\newcommand{\twosevensixnn}{$\sqrt{s_{\mathrm{NN}}}~=~2.76$~Te\kern-.1emV\xspace}
\newcommand{\fivenn}       {$\sqrt{s_{\text{\tiny NN}}}=5.02$~Te\kern-.1emV\xspace}
\newcommand{\twoHnn}       {$\sqrt{s_{\text{\tiny NN}}}=200$~Ge\kern-.1emV\xspace}
\newcommand{\LT}           {L{\'e}vy-Tsallis\xspace}
\newcommand{\GeVc}         {Ge\kern-.1emV/$c$\xspace}
\newcommand{\MeVc}         {Me\kern-.1emV/$c$\xspace}
\newcommand{\TeV}          {Te\kern-.1emV\xspace}
\newcommand{\GeV}          {Ge\kern-.1emV\xspace}
\newcommand{\MeV}          {Me\kern-.1emV\xspace}
\newcommand{\GeVmass}      {Ge\kern-.2emV/$c^2$\xspace}
\newcommand{\MeVmass}      {Me\kern-.2emV/$c^2$\xspace}
\newcommand{\lumi}         {\ensuremath{\mathcal{L}}\xspace}

\newcommand{\ITS}          {\rm{\ac{ITS}}\xspace}
\newcommand{\TOF}          {\rm{\ac{TOF}}\xspace}
\newcommand{\ZDC}          {\rm{\ac{ZDC}}\xspace}
\newcommand{\ZDCs}         {\rm{ZDCs}\xspace}
\newcommand{\ZNA}          {\rm{ZNA}\xspace}
\newcommand{\ZNC}          {\rm{ZNC}\xspace}
\newcommand{\SPD}          {\rm{\ac{SPD}}\xspace}
\newcommand{\SDD}          {\rm{\ac{SDD}}\xspace}
\newcommand{\SSD}          {\rm{\ac{SSD}}\xspace}
\newcommand{\TPC}          {\rm{\ac{TPC}}\xspace}
\newcommand{\TRD}          {\rm{\ac{TRD}}\xspace}
\newcommand{\VZERO}        {\rm{V$0$}\xspace}
\newcommand{\TZERO}        {\rm{T$0$}\xspace}
\newcommand{\TZEROA}        {\rm{T$0$A}\xspace}
\newcommand{\TZEROC}        {\rm{T$0$C}\xspace}
\newcommand{\VZEROA}       {\rm{V$0$A}\xspace}
\newcommand{\VZEROC}       {\rm{V$0$C}\xspace}
\newcommand{\Vdecay} 	   {\ensuremath{V^{0}}\xspace}

\newcommand{\ee}           {\ensuremath{e^{+}e^{-}}} 
\newcommand{\pip}          {\ensuremath{\pi^{+}}\xspace}
\newcommand{\pim}          {\ensuremath{\pi^{-}}\xspace}
\newcommand{\piz}          {\ensuremath{\pi^{0}}\xspace}
\newcommand{\pipipi}       {\ensuremath{\pi^{+}\pi^{-}\pi^{0}}\xspace}
\newcommand{\kap}          {\ensuremath{\rm{K}^{+}}\xspace}
\newcommand{\kam}          {\ensuremath{\rm{K}^{-}}\xspace}
\newcommand{\pbar}         {\ensuremath{\rm\overline{p}}\xspace}
\newcommand{\kzero}        {\ensuremath{{\rm K}^{0}_{\rm{S}}}\xspace}
\newcommand{\lmb}          {\ensuremath{\Lambda}\xspace}
\newcommand{\almb}         {\ensuremath{\overline{\Lambda}}\xspace}
\newcommand{\Om}           {\ensuremath{\Omega^-}\xspace}
\newcommand{\Mo}           {\ensuremath{\overline{\Omega}^+}\xspace}
\newcommand{\X}            {\ensuremath{\Xi^-}\xspace}
\newcommand{\Ix}           {\ensuremath{\overline{\Xi}^+}\xspace}
\newcommand{\Xis}          {\ensuremath{\Xi^{\pm}}\xspace}
\newcommand{\Oms}          {\ensuremath{\Omega^{\pm}}\xspace}
\newcommand{\degree}       {\ensuremath{^{\rm o}}\xspace}

\newcommand{\reffig}[1]{\hyperref[fig:#1]{Fig.~\ref*{fig:#1}}}
\newcommand{\refsec}[1]{\hyperref[sec:#1]{Sec.~\ref*{sec:#1}}}
\newcommand{\reftab}[1]{\hyperref[tab:#1]{Tab.~\ref*{tab:#1}}}
\newcommand{\refFig}[1]{\hyperref[fig:#1]{Figure~\ref*{fig:#1}}}
\newcommand{\refSec}[1]{\hyperref[sec:#1]{Section~\ref*{sec:#1}}}
\newcommand{\refTab}[1]{\hyperref[tab:#1]{Table~\ref*{tab:#1}}}

\newcommand{\Pythia}          {PYTHIA~8\xspace}
\begin{acronym}[UMLX]
    \acro{ALICE}{A Large Ion Collider Experiment}
    \acro{ATLAS}{A Toroidal LHC ApparatuS}
    \acro{BFKL}{Balitsky-Fadin-Kuraev-Lipatov}
    \acro{BNL}{Brookhaven National Laboratory}
    \acro{CERN}{Conseil Européen pour la Recherche Nucléaire}
    \acro{CGC}{Color Glass Condensate}
    \acro{CMS}{Compact Muon Solenoid}
    \acro{CNM}{Cold Nuclear Matter}
    \acro{d--Au}{Deutrerium--Gold}
    \acro{DCA}{Distance of Closest Approach}
    \acro{DCal}{Di-Jet Calorimeter}
    \acro{DPM}{Dual Parton Model}
    \acro{GEANT}{GEometry ANd Tracking}
    \acro{DGLAP}{Dokshitzer-Gribov-Lipatov-Alterelli-Parisi}
    \acro{ECRIS}{Electron Cyclotron Resonance Ion Source}
    \acro{EMC}{European Muon Collaboration}
    \acro{EMCal}{Electromagnetic Calorimeter}
    \acro{FAIR}{Facility for Antiproton and Ion Research}
    \acro{FWHM}{Full Width at Half Maximum}
    \acro{FF}{Fragmentation Function}
    \acro{ISR}{Intersecting Storage Rings}
    \acro{ITS}{Inner Tracking System}
    \acro{LHC}{Large Hadron Collider}
    \acro{LHCb}{LHC beauty}
    \acro{LEIR}{Low Energy Ion Ring}
    \acro{LINAC}{Linear Accelerator}
    \acro{MB}{Minimum Bias}
    \acro{MBW}{Material Budget Weight}
    \acro{MC}{Monte-Carlo}
    \acro{MIP}{Minimum-Ionizing Particle}
    \acro{MRPC}{Multigap Resistive Plate Chamber}
    \acro{MWPC}{Multi Wire Proportional Chamber}
    \acro{NSD}{Non Single Diffractive}
    \acro{PDG}{Particle Data Group}
    \acro{PHOS}{Photon Spectrometer}
    \acro{QGP}{Quark Gluon Plasma}
\acro{RHIC}{Relativistic Heavy Ion Collider}
\acro{SDD}{Silicon Drift Detector}
    \acro{SPD}{Silicon Pixel Detector}
    \acro{SPS}{Super Proton Synchrotron}
    \acro{SSD}{Silicon Strip Detector}
    \acro{TPC}{Time Projection Chamber}
    \acro{TRD}{Transition Radiation Detector}
    \acro{TOF}{Time Of Flight}
    \acro{pp}{Proton-Proton}
    \acro{p--Pb}{Proton--Lead}
    \acro{Pb--Pb}{Lead--Lead}
    \acro{PS}{Proton Synchrotron}
    \acro{PSB}{Proton Synchrotron Booster}
    \acro{pQCD}{perturbative \acs{QCD}}
    \acro{PDF}{Parton Distribution Function}
    \acro{nPDF}{nuclear PDF}
    \acro{BLUE}{Best Linear Unbiased Estimator}
    \acro{ESD}{Simulation}
    \acro{PCM}{Photon Conversion Method}
    \acro{PID}{Particle Identification}
    \acro{PHENIX}{Pioneering High Energy Nuclear Interaction eXperiment}
    \acro{GSI}{Gesellschaft für Schwerionenforschung}
    \acro{RMS}{Root Mean Square Value}
    \acro{TCM}{Two Component Model}
    \acro{QA}{Quality Assurance}
    \acro{QFT}{Quantum Field Theory}
    \acro{QCD}{Quantum Chromodynamics}
\end{acronym}
\begin{titlepage}
\PHyear{2025}       
\PHnumber{033}      
\PHdate{24 February}  

\makeatletter
\newcommand*{\link}{\begingroup\@makeother\#\@mylink}
\newcommand*{\@mylink}[2]{\href{#1}{\underline{#2}}\endgroup} 
\makeatother

\title{Measurement of $\omega$ meson production\\in pp and p--Pb collisions at $\sqrt{s_{\text{\footnotesize NN}}}$~=~5.02~Te\kern-.1emV}
\ShortTitle{Production of $\omega$ mesons in pp and p--Pb collisions at \fivenn}   

\Collaboration{ALICE Collaboration\thanks{See Appendix~\ref{app:collab} for the list of collaboration members}}
\ShortAuthor{ALICE Collaboration} 

\begin{abstract}
We present the measurement of the \pt-differential production cross section of $\omega$ mesons in pp and p--Pb collisions at \fivenn at midrapidity by ALICE. In addition, the first measurement of the nuclear modification factor $R_{\text{pPb}}$ for $\omega$ mesons at LHC energies is presented, complementing the existing measurements of lighter neutral mesons such as the \piz and $\eta$. Within the measured \pt-range, the $R_{\text{pPb}}$ of $\omega$ mesons is compatible with no cold nuclear matter effects within the uncertainties, consistent with previous measurements at lower energies. The $\omega/\pi^0$ ratio is presented for both collision systems, showing no collision system dependence within the uncertainties. The comparison to previously published $\omega/\pi^0$ ratios at lower and higher collision energies in pp collisions suggests a decreasing trend of the ratio above \pt~=~4~\GeVc with increasing collision energy.
The data in both collision systems are compared to predictions from \Pythia, EPOS LHC and DPMJET event generators, revealing significant shortcomings in these models' ability to describe the production of $\omega$ mesons.
\end{abstract}
\end{titlepage}

\setcounter{page}{2} 


\section{Introduction} 
\label{sec:Introduction}
Measurements of hadron production in ultra-relativistic nuclear collisions have substantially improved the understanding of \ac{QCD}, the theory of the strong interaction~\cite{Jacob:156665,ALICE:2022wpn}. 
While pp collisions serve as a vacuum baseline, analyses of p--A and A--A collisions reveal emergent \ac{QCD} phenomena, such as collective flow and energy loss in cold and hot nuclear matter~\cite{ALICE:2022wpn}.
Theoretically, particle production in such collisions can be divided into the hard and soft regimes, denoting scattering processes with large and small momentum transfer $Q^{2}$, respectively.
Hard processes, if not modified by energy loss in the \ac{QCD} medium, can be calculated using perturbative \ac{QCD}, however, this relies on experimental input for the \acp{PDF} and \acp{FF}~\cite{nCTEQnPDFs,FFs,Saveetha_2017}. 
On the other hand, soft processes are not calculable perturbatively, and phenomenological models and \ac{MC} event generators are used to describe them.

Studying particle production in collisions of protons and heavy nuclei, such as \pPb collisions, gives insights into the influence of \ac{CNM}.
Modifications of particle production with respect to the vacuum baseline, namely pp collisions, can be estimated using the nuclear modification factor, which, in the case of \pPb collisions, is defined as
\begin{equation}
    R_{\text{pPb}} = \frac{1}{A_{\text{Pb}}} \left(E\frac{\text{d}^{3}\sigma_{\text{pPb}}}{\text{d}p^{3}}\right)  \Big/ \left(E\frac{\text{d}^{3}\sigma_{\text{pp}}}{\text{d}p^{3}}\right),
\end{equation}
where $E\frac{\text{d}^{3}\sigma_{\text{pPb (pp)}}}{\text{d}p^{3}}$ denotes the measured production cross section of a certain particle species in p--Pb  (pp) collisions, and $A_\text{Pb} = 208$ represents the number of nucleons in a lead nucleus. 
It was found in measurements at the \ac{RHIC} and the \ac{LHC}, that the nuclear modification factor is in agreement with unity above \pt~$\approx$~2~\GeVc for various mesons and baryons~\cite{ALICE:2018vhm, 2016313, ALICE:2012mj, ALICE:2021est, PhysRevLett.98.172302, Adams_2006, CMS:2014RpA, ALICE:2013wgn, ALICE:2016dei}. At lower transverse momenta (\pt~$<$~2~\GeVc), a significant depletion is observed, showcasing the presence of CNM effects~\cite{ALICE:2018vhm, 2016313, ALICE:2012mj}. 

There are different approaches to explain the modification theoretically:
nuclear \ac{PDF}s (n\ac{PDF}s) are used to model the \ac{PDF} for single nucleons in a nucleus. 
Since the first observations of modification of the \ac{PDF} in such environments~\cite{EMCEffect, ARNEODO1988493}, the understanding of those nPDFs has been refined, leading to very precise knowledge of their evolution with Bjorken $x$ and momentum transfer $Q$~\cite{Klasen_2024}.
Nuclear shadowing, a depletion of the \ac{PDF} when measured for a nucleon inside a nucleus for $x \lessapprox 0.01$, is the cause for a reduction of the observed production cross section of particles below \pt~$\approx$~3~\GeVc. Furthermore, a slight enhancement of the \ac{PDF} is predicted at $x \approx 0.1$, known as anti-shadowing.

In the \ac{CGC} model, suppression arises from high parton densities, resulting in gluon saturation in the low-\pt regime~\cite{Gelis_2010}. Calculations using the \ac{CGC} successfully describe the suppression of particle production at low transverse momenta. 
High-\pt measurements at midrapidity, however, probe a region far from the saturation scale, for which those calculations are not applicable. 

An alternative to modifications of the parton distributions in the nucleus is to explain the depletion of particle production at low \pt in \pPb collisions via fully coherent energy loss (FCEL) induced by gluon radiation in the \ac{CNM} \cite{FCEL}.

While measurements of light mesons and baryons are abundant, measurements of heavier particles, like the $\omega$ meson, are scarce, mostly due to more complicated decay chains compared to lighter particles, like the neutral pion. However, measurements of these heavier particles give insights into the mass dependence of particle production and the nuclear modification factor. In the case of the $\omega$ meson, the comparison to the neutral pion also reveals a possible spin dependence (spin 1 and spin 0, respectively), as the quark content of the \piz and $\omega$ is almost the same~\cite{PDG}.

This paper is structured as follows:
first, a brief overview of the ALICE experiment and the detector systems involved in the analysis will be presented in \refsec{Detector}, followed by an overview of the datasets in \refsec{Datasets}. 
The reconstruction of charged pions, photons and neutral pions is given in \refsec{Pions}.
The $\omega$ meson reconstruction is described in section \refsec{Omega}, followed by a discussion of the systematic uncertainties in \refsec{Systematics}. 
Finally, the results are being discussed in \refsec{Results}, and the conclusion is presented in \refsec{Conclusion}.
 
\section{ALICE detector} 
\label{sec:Detector}
The $\omega$ mesons are reconstructed via their decay into three pions and the subsequent decay of the neutral pion into a pair of photons: $\omega\rightarrow\pipipi\rightarrow\pi^+\pi^-\gamma\gamma$. 
The charged pions are identified using the \ac{TPC} and \ac{TOF} detector information of the associated tracks reconstructed in \acs{ALICE}'s central tracking detectors, the \ac{ITS} and the \ac{TPC}.
The photons are measured in the \ac{EMCal}, or, in the case the photon converts into a dielectron pair in the detector material, they are reconstructed from their $e^+e^-$ tracks in the tracking detectors.
The following paragraphs briefly describe the five key detectors utilized in the analysis, in their configuration during \acs{LHC} Run 2 (2015-2018). Comprehensive details on the ALICE detectors and their performance are available in Refs.~\cite{ALICE_Overview} and \cite{ALICE:2014sbx}.

The \ac{ITS}~\cite{ITS_TDR, ALICE:2010tia} consists of six cylindrical layers of silicon detectors surrounding the beam pipe, going from the innermost to the outermost layer: two layers of Silicon Pixel Detectors (SPD), two layers of Silicon Drift Detectors (SDD), and two layers of Silicon Strip Detectors (SSD). This analysis utilizes the SPD and SSD, covering the full azimuthal angle and a pseudorapidity range of $\vert \eta \vert <$~0.9, to locate the collision vertex and to provide the innermost clusters for tracking charged particles. 

The \ac{TPC}~\cite{TPCTDR} is a cylindrical drift detector surrounding the ITS. Charged particles ionize the gas atoms along their trajectory within a pseudorapidity coverage of $\vert \eta \vert < 0.9$ and full azimuthal coverage. 
Due to an applied electric field, the freed electrons then drift to the electrodes on either end of the TPC, multiplied using Multi Wire Proportional Chambers, and finally the signal is read out using cathode pads~\cite{TPC_TechNote}.

The \VZERO detector~\cite{ALICE:2013axi} is made up of two scintillation counters \VZEROA and \VZEROC, placed on either side of the interaction point along the beam axis, covering the full azimuthal angle for -3.7~$<\eta<$~-1.7 as well as 2.8~$<\eta<$~5.1, respectively. The \VZERO detector measures the event multiplicity in the forward region and also serves as ALICE's \ac{MB} trigger detector. The adjacent \TZERO detectors \TZEROA and \TZEROC~\cite{T0} consist of twelve Cherenkov detectors, covering the full azimuthal angle for -3.3~$<\eta<$~-2.9 and 4.5~$<\eta<$~5.0. With a time resolution of $50~$ps, the \TZERO detectors are used to determine the collision time as a reference for other detectors, such as the \ac{EMCal}, \ac{TPC} and \ac{TOF}.

The Time Of Flight (TOF) detector~\cite{TOF_TDR} surrounds the beam pipe with an array of Multigap Resistive Plate Chamber strips at a distance of about 3.7\,m, thereby covering the full azimuthal angle within $\vert\eta\vert$~$<$~0.9. In combination with the collision time signal provided by the \TZERO detectors, the TOF measures the flight time of charged particles.

The \ac{EMCal}~\cite{EMCal_TDR} is a sampling calorimeter with alternating layers of lead and scintillation material. Within ${\vert \eta \vert < 0.7}$, it covers the azimuthal angles between 80~$<\varphi$ $({}^{\circ}) < $~187 and 261~$ < \varphi$ $({}^{\circ}) < $~319. Photons can be reconstructed using the \ac{EMCal} from their electromagnetic showers, starting with a minimum energy of $E_\text{min}=0.7~$\GeV~\cite{ALICE:2022qhn}.
\section{Datasets and event selection} 
\label{sec:Datasets}
The analysis utilizes all datasets of \pp and \pPb collisions at \fivenn recorded by ALICE in the second run of the LHC. The collisions used for this analysis fulfill the minimum bias (MB) trigger condition of a coincidental signal in both arms of the \VZERO detector.
This includes approximately one billion \pp collisions recorded in November 2015 and 2017 and around half as many \pPb collisions collected in late November 2016. These datasets correspond to an inspected integrated luminosity in \pp (\pPb) collisions of $\mathcal{L}_\text{int}=N_\text{evt}/\sigma_\text{MB}=18~$n$\text{b}^{-1}$ ($0.27~$n$\text{b}^{-1}$).
The respective visible minimum bias cross sections $\sigma_\text{MB}$ were determined via Van der Meer scans~\cite{vanderMeer:296752} for \fivenn in \pp~\cite{ALICE:2018lum} and \pPb~\cite{ALICE:2014gvw} collisions.

The primary vertex of each collision is reconstructed via tracks in the TPC and ITS, as described in Ref.~\cite{ALICE:2014sbx}. 
The distance in the beam direction (z) between the vertex and the nominal interaction point is required to be within $\vert z_\text{vtx}\vert$\,$<$\,10~cm, to ensure good vertex resolution and uniform detector acceptance. 
To account for triggered collisions in which no vertex is reconstructed, the luminosity used for the normalization was corrected upward by 1.4\% (0.5\%) in \pp (\pPb) collisions, assuming no $\omega$ mesons are produced at midrapidity in such events.
Pileup collisions with more than one collision vertex reconstructed in the SPD in the same bunch crossing are rejected; this is the case for 0.2\% (3.7\%) of \pp (\pPb) collisions.

In \pPb collisions, the nucleon-nucleon center-of-mass system is boosted with $\Delta y_{\rm NN}=0.465$ in the direction of the proton beam due to the 2-in-1 magnet design of the LHC~\cite{LHCMachine}, which uses the same magnetic field for both beams, accelerating the protons to higher velocities because of their larger charge-to-mass ratio.
\section{Pion reconstruction} 
\label{sec:Pions}
As already mentioned, in the presented measurement, $\omega$ mesons are reconstructed via their dominant decay into $\pi^{+}\pi^{-}\pi^{0}$ ($\mathcal{B}$~$\approx$~89.2\%~\cite{PDG}).  
This section discusses the reconstruction of the charged pions, followed by the reconstruction of the neutral pions. The selection criteria and reconstruction methods are the same for the measurement in \pp and \pPb collisions.

\subsection{Charged pion selection}

Charged particles can be reconstructed from clusters along their trajectory using a Kalman filter, with their momentum measured through their curvature in ALICE's magnetic field; see Ref.~\cite{ALICE:2014sbx} for details.
With a lifetime of $\tau^{\pi^\pm}=7.8$~m/$c$~\cite{PDG}, the vast majority of charged pions traverse the tracking detectors, \ITS and \TPC, so they can be reconstructed from their tracks.
To ensure good quality of the tracks, they are required to have a transverse momentum of $\pt>$\,100~\MeVc, contain at least one hit in the SPD, include at least 80 TPC clusters and $\chi^2/N_\text{cluster}^\text{TPC}$~$<$~4.

The charged pions are identified based on each track's specific energy loss \dEdx in the TPC, quantified via the deviation of the measured energy loss from the expected energy loss relative to the detector resolution. 
This relative deviation is required to be $\vert n\sigma^{\pi^\pm}_\text{TPC}\vert<$\,3.
For charged tracks that include a TOF cluster, the relative deviation from the expected flight time for pions relative to the detector resolution is examined to increase the purity of the pion selection further.
Thus, charged particles that the TOF classified as unlikely to be a charged pion ($\vert n\sigma^{\pi^\pm}_\text{TOF}\vert>5$) and likely to be either a proton or a kaon ($\vert n\sigma^{p/K}_\text{TOF}\vert<$\,3) were removed from the sample of selected pion tracks.

With the selection criteria described above, the reconstruction efficiency of charged pions is around $\epsilon^{\pi^\pm} \approx$\,80\% for $\pt>500~$MeV, with a purity of $P^{\pi^\pm}$\,=\,99\% at $\pt\approx500~$MeV/$c$, decreasing to $P^{\pi^\pm}$\,=\,80\% at $\pt\approx7~$GeV/$c$.

\subsection{Neutral pion reconstruction}

Neutral pions decay at the primary vertex with a branching ratio of $\mathcal{B}$~$\approx$~99.8\%~\cite{PDG} into a photon pair. 
The decay photons are reconstructed using either the EMCal or the Photon Conversion Method (PCM), as laid out below.

Photons reaching the EMCal produce electromagnetic showers that typically deposit their energy in multiple nearby cells. 
Clusters are formed by grouping adjacent cells, starting with seed cells that have an energy $E_\text{seed}>500~$MeV and then aggregating adjacent cells with an energy $E_\text{agg}>100~$MeV, as outlined in Ref.~\cite{ALICE:2022qhn}.
The cluster energy $E_\text{cls}$ is obtained as the sum of the individual cell energies, with an additional correction to account for the non-linearity of the energy response~\cite{ALICE:2022qhn}.
Furthermore, the signal time of each cluster relative to the collision time has to fulfill -20\,$\leq\,t_\text{cls}~$(ns)\,$<$\,25 to reduce out-of-bunch pileup.
Only clusters with an energy of $E_\text{cls} > 0.7~$GeV are considered in the analysis to reduce the number of clusters induced by minimum ionizing particles.
The shape of clusters is characterized by $\sigma_\text{long}^2$, quantifying the elongation of the underlying electromagnetic shower; for more details, see Ref.~\cite{ALICE:2022qhn}.
The elongation of clusters containing more than one cell must be within 0.1\,$< \sigma_\text{long}^2 <$\,0.7 to reduce contributions from the merged clusters of $\piz$ decay photons and those from the low-\pt electrons and hadrons.
In a final selection step, clusters are rejected if a charged particle track, reconstructed in TPC and ITS, is found within a \pt-dependent distance in $\eta$ and $\varphi$ from the cluster, as described in Ref.~\cite{ALICE:2017ryd}.

Approximately 8.5\%~\cite{ALICE:2014sbx} of photons passing through the ALICE inner detector material convert into an $e^+ e^-$ pair within a radial distance of $180~$cm from the beam axis. 
The first step in reconstructing these photons is to select the electron and positron tracks measured in the central tracking detectors. 
Tracks are required to have a transverse momentum of $\pt>40$~\MeVc and to contain at least 60\% of the TPC clusters that are expected from the track's trajectory.
The track's measured energy loss \dEdx in the TPC is required to be close to that expected for electrons (-3\,$<n\sigma^\text{e}_\text{TPC}<$\,4). 
In order to reduce contamination from charged pions, the measured energy loss is furthermore required to be more than one sigma away from the expected energy loss of pions ($\vert n\sigma^{\pi^\pm}_\text{TPC}\vert>$\,1) for transverse momenta above $\pt>0.4~$\GeVc.
The distinct V-shaped topology of two tracks with opposite curvatures originating from a common secondary vertex is identified with a $\text{V}^0$ finding algorithm~\cite{ALICE:2014sbx}. 
The distance of the conversion point from the beam axis is required to be $R>5~$cm to reduce contamination from Dalitz decays but within $R<180~$cm to ensure good track quality. 
Further selections that enhance the purity of conversion photons in this $\text{V}^0$ candidate sample, such as requiring the momentum vector of the $\text{V}^0$ to point to the primary vertex, have been applied as described in Ref.~\cite{ALICE:13TeVpi0}.

These two photon-reconstruction methods provide three methods of reconstructing neutral pions. These are called PCM or EMCal if both of the decay photons are reconstructed using the same method, or PCM-EMCal, if one photon is measured in the calorimeter while the other is reconstructed with PCM.
Neutral pion candidates are accepted if the invariant mass of the photon pair is within $M^{\piz}_\text{rec} (\pt) \pm~n\sigma^{\piz}_\text{rec} (\pt)$, where $M^{\piz}_\text{rec}$ and $\sigma^{\piz}_\text{rec}$ are the \pt-dependent mass and width of the reconstructed neutral pion taken from Ref.~\cite{ALICE:13TeVpi0}.
The parameter $n$ is chosen to be $3$ for PCM, $2.5$ for PCM-EMCal, and $2$ for EMCal to increase the available statistics for the PCM analysis while prioritizing the enhanced purity for the EMCal method. 
\section{Reconstruction of \texorpdfstring{$\omega$}{omega} mesons}
\label{sec:Omega}

The transverse momentum and invariant mass of $\omega$ meson candidates is calculated from the four-momenta of all possible \pipipi combinations of selected pions in each event.
As the $\omega$ meson decays via the strong interaction, with a lifetime of $\tau^\omega$~$\approx$~23.2~fm/$c$~\cite{PDG}, all decay products are assumed to originate from the collision vertex.

The energy resolution of the involved detectors causes a smearing of the reconstructed mass of the $\omega$ meson candidates, thereby reducing the signal-to-background ratio in an invariant-mass-based signal extraction. The dominant fraction of the smearing comes from the reconstruction of the $\pi^0$, whose mass is similarly affected by the detector resolution.
As demonstrated in previous publications~\cite{ALICE:2020ylo, ALICE:13TeVomega}, this effect can be mitigated by shifting the reconstructed invariant mass of an $\omega$ candidate by the difference of the reconstructed \piz invariant mass to the literature value of $M^{\piz}_\text{PDG}$~=~134.98~MeV/$c^2$~\cite{PDG}.In the presented measurement, a novel approach is introduced to further refine this correction by determining the correlation between a given discrepancy of the reconstructed \piz mass ($M^{\piz}$) and the corresponding discrepancy of the reconstructed $\omega$ mass ($M^\omega$), quantified by the ratio
$\lambda = \langle \left(M^\omega_\text{rec}-M^\omega_\text{gen}\right)/\left(M^{\piz}_\text{rec}-M^{\piz}_\text{PDG}\right) \rangle$, using Monte-Carlo simulations.
Here, $M^\omega_\text{gen}$ corresponds to the reference mass that the $\omega$ meson was generated with and is extracted from the simulation. This correlation $\lambda$ was found to depend on the opening angle between the decay photons of the \piz due to the interplay of the photon energy resolution and the resulting position resolution of the neutral pion candidate. Depending on the reconstruction method and opening angle, its value ranges between $1.3<\lambda<2.3$. For further details and figures, see the supplemental note~\cite{ALICE:SuppNote}.
This correlation is then used to shift the reconstructed mass of each $\pipipi$ combination individually based on the deviation of the contained $\piz$ from the PDG mass: $M^{\pipipi} = M^{\pipipi}_\text{rec} - \lambda \left( M^{\piz}_\text{rec} - M^{\piz}_\text{PDG} \right)$.

\begin{figure}
    \centering
    \includegraphics[width = 0.495\textwidth]{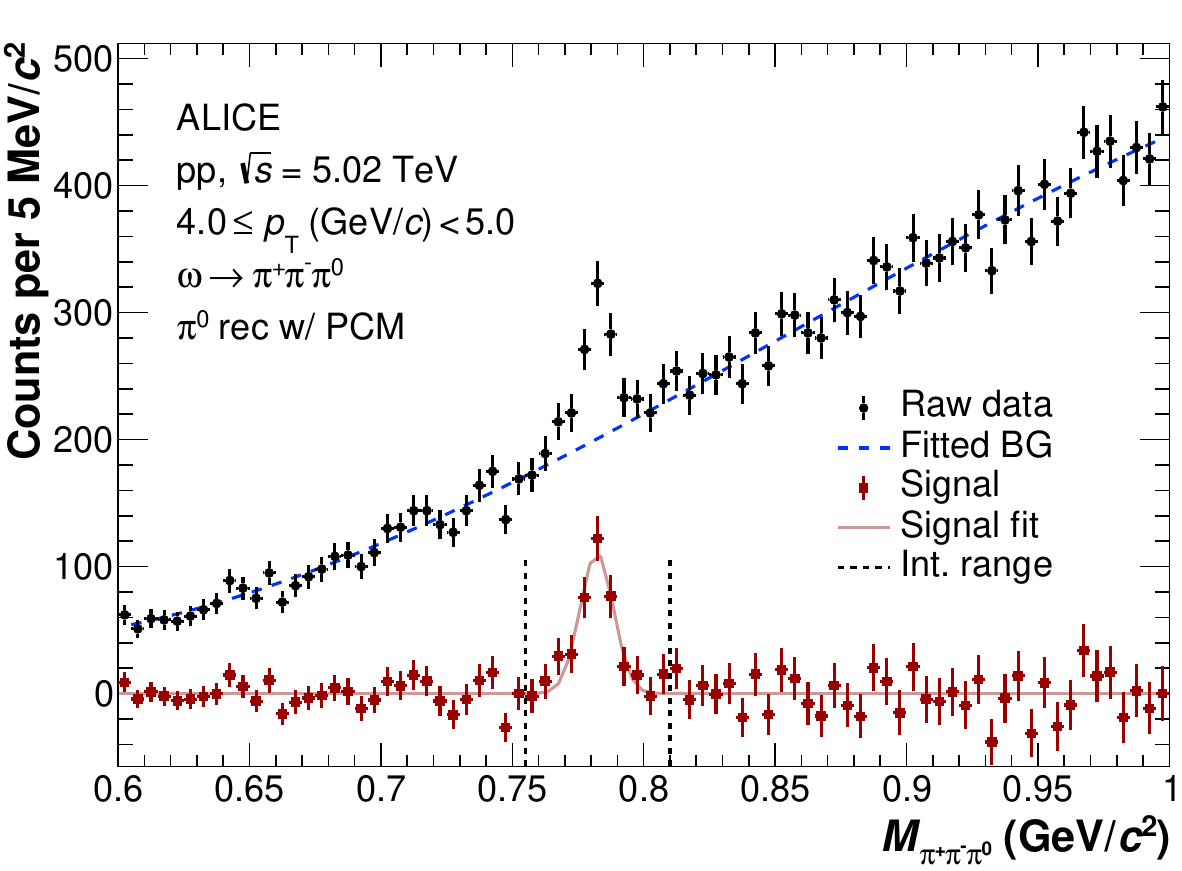}
    \includegraphics[width = 0.495\textwidth]{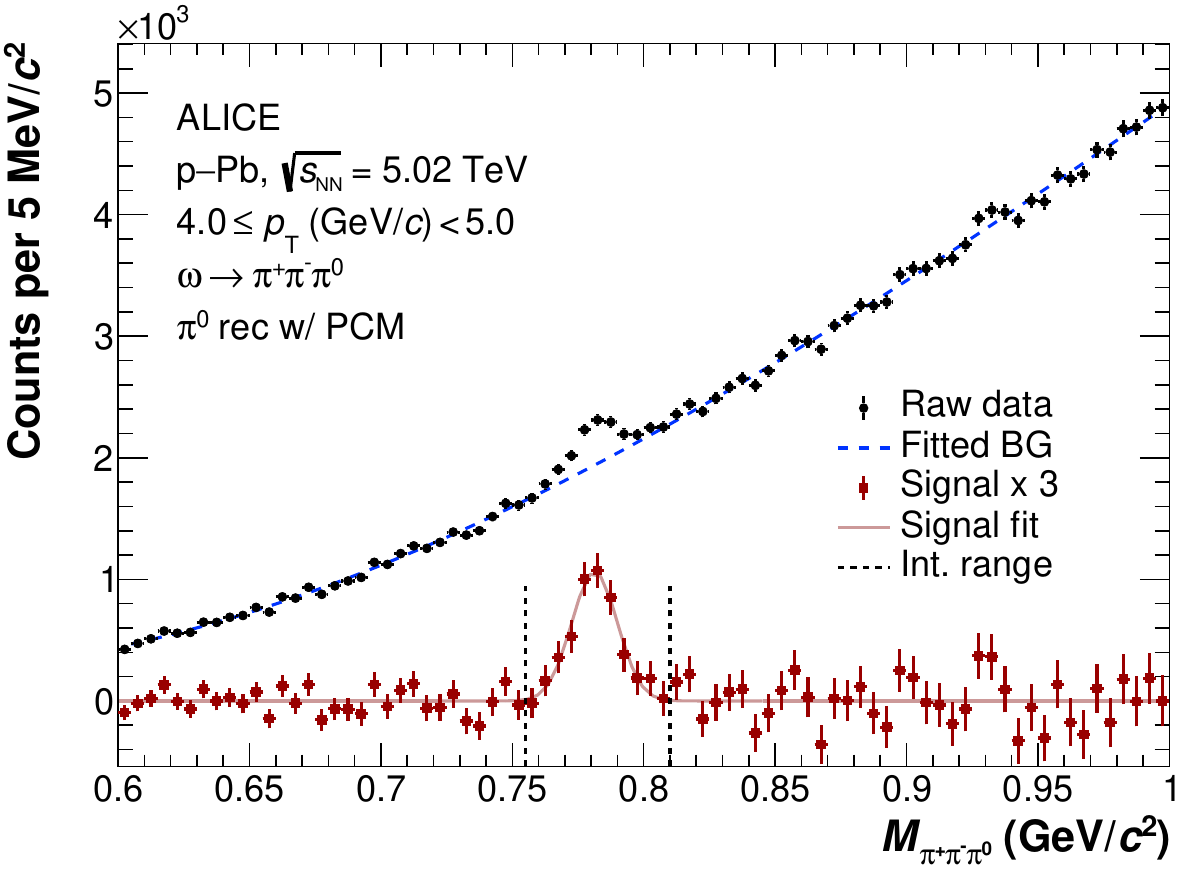}
    \includegraphics[width = 0.495\textwidth]{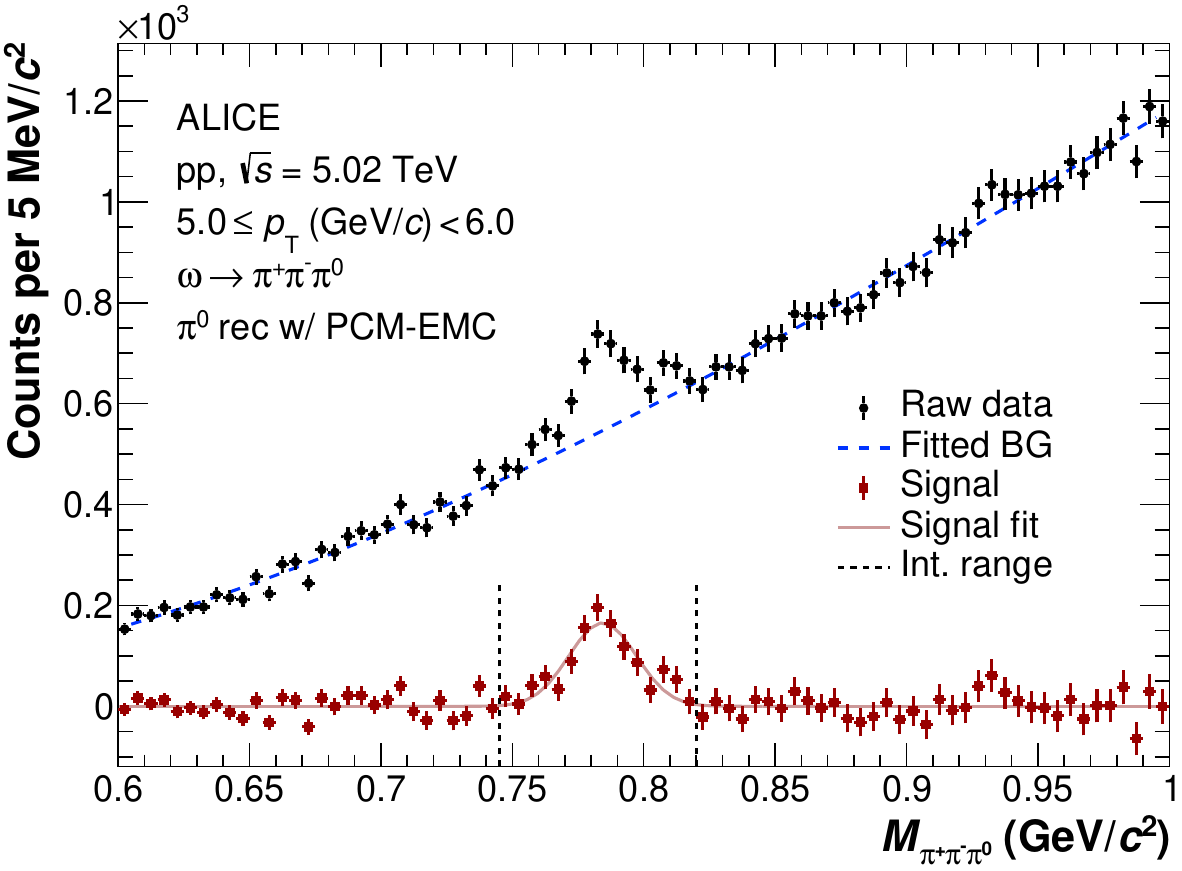}
    \includegraphics[width = 0.495\textwidth]{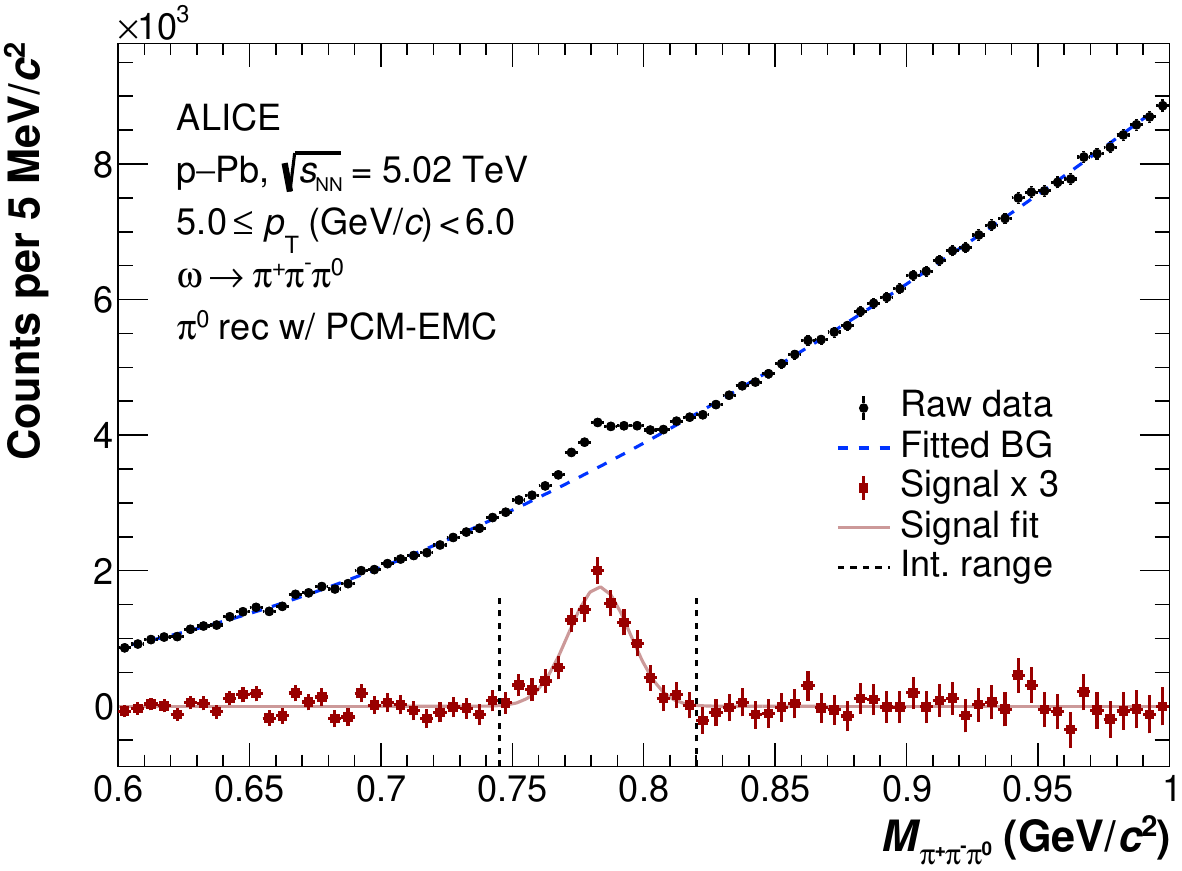}
    \includegraphics[width = 0.495\textwidth]{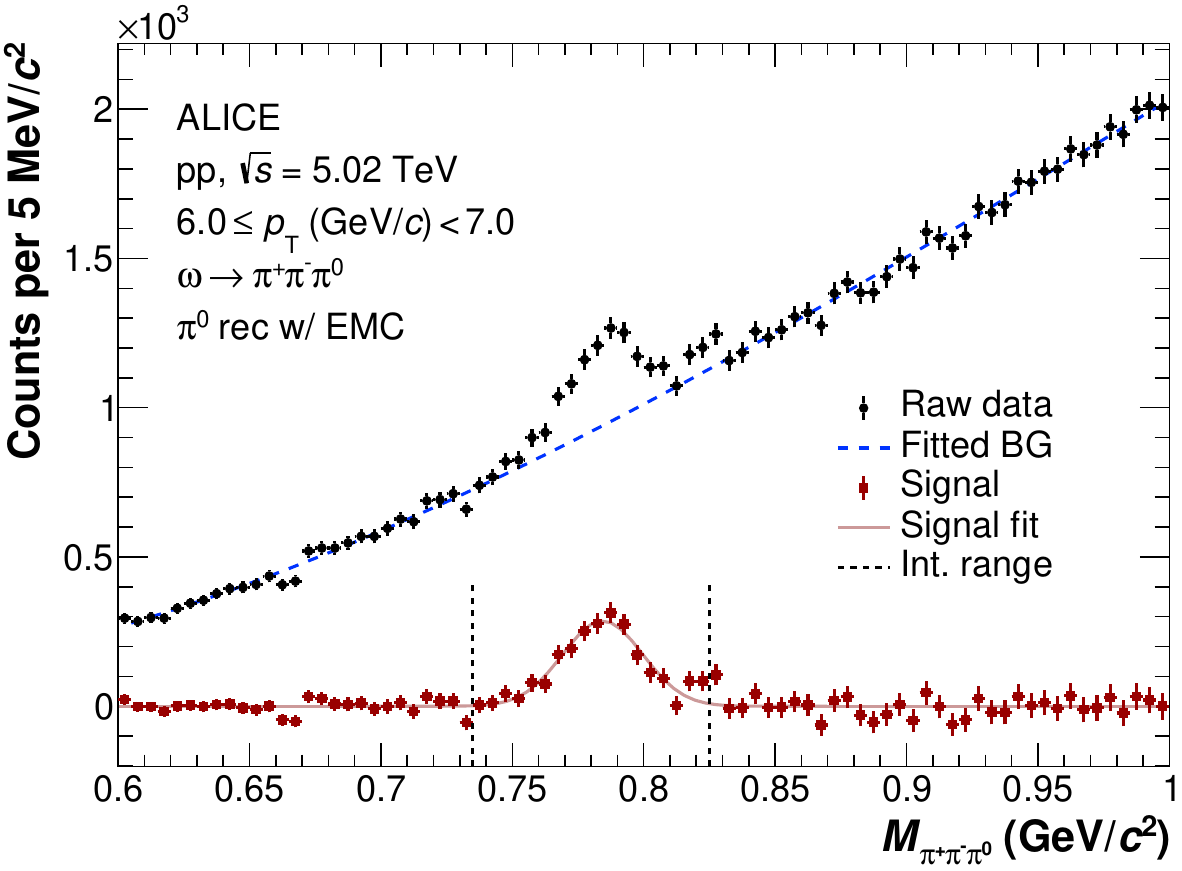}
    \includegraphics[width = 0.495\textwidth]{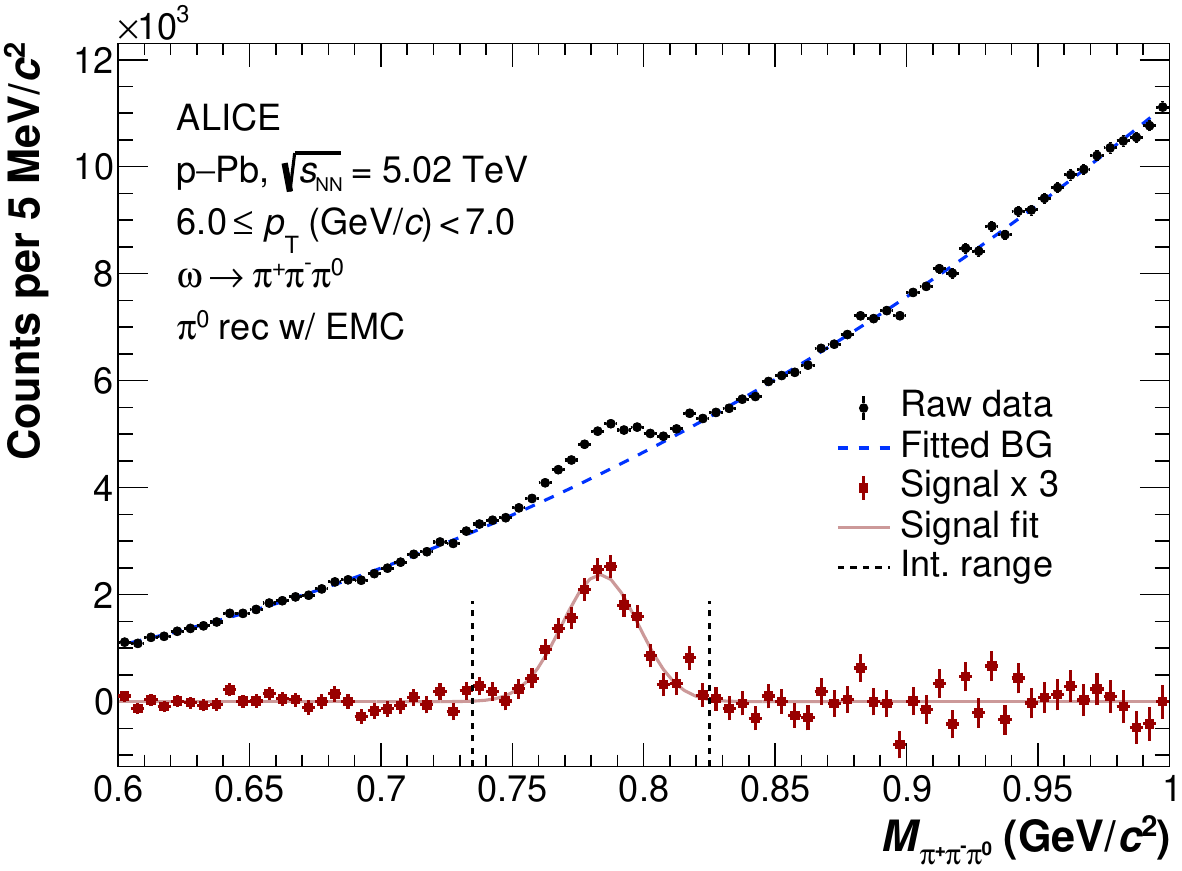}
    \caption{Invariant mass distributions of \pipipi triplets in the indicated \pt ranges for \piz's reconstructed with PCM (top), PCM-EMCal (middle), and EMCal (bottom) in \pp (left) and \pPb (right) collisions. Subtraction of the third-order polynomial background description (blue) from the \pipipi candidates (black) results in the signal of $\omega$ candidates (red). In \pPb collisions, this signal is scaled by a factor of three for better visibility. The vertical dashed lines denote the invariant mass region where the raw yield was obtained through bin counting, as described in \refsec{Omega}.}
    \label{fig:ExampleBins}
\end{figure}

The obtained invariant-mass distributions for one \pt interval for each \piz reconstruction method and collision system are shown in~\reffig{ExampleBins} as black markers. 
A signal peak around the nominal mass of the $\omega$ meson is visible on top of the combinatorial background. 
The background is described using a third-order polynomial parametrization, shown as a blue line in~\reffig{ExampleBins}. It is obtained from a fit of the invariant mass distribution excluding a signal region of $M^\omega \pm 3\sigma^\omega$.
This signal region is defined using a preliminary, \pt-independent mass and width, extracted via Gaussian parametrizations of the invariant mass distributions for each reconstruction method.
The signal invariant-mass distribution, shown in red, is extracted by subtracting the parametrized combinatorial background from the invariant mass distribution.  
The signal-to-background ratio at low \pt is approximately three times higher in \pp collisions than in \pPb collisions, which coincides with an approximately three times larger event multiplicity in the latter~\cite{ALICE:2022xip}.
Due to the better energy resolution of PCM compared to the EMCal, the width of the peaks, and thereby the signal-to-background ratio, also vary between the reconstruction methods.
The raw $\omega$ yield for each \pt interval is then calculated as the sum of bin contents within the aforementioned signal region, denoted by black dashed lines in \reffig{ExampleBins}.

\begin{figure}[tb]
    \centering
    \includegraphics[width = 0.495\textwidth]{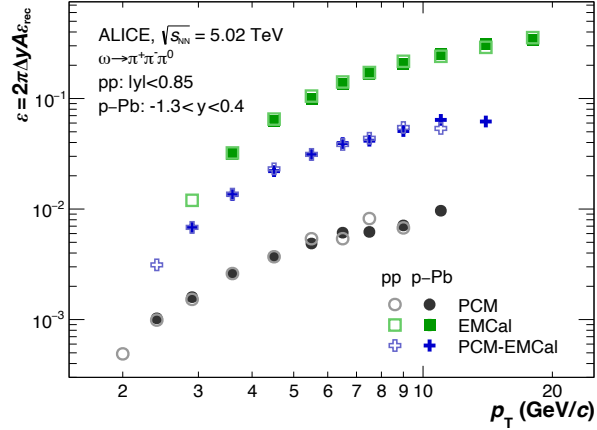}
    \caption{Correction factors $\epsilon$ applied to the raw $\omega$ yields for each \piz reconstruction method in \pp and \pPb collisions. The correction factor comprises the geometrical acceptance ($A$), the reconstruction efficiency ($\epsilon_\text{rec}$), and normalization to azimuthal and rapidity ranges.}
    \label{fig:Corrections}
\end{figure}

The reconstruction efficiency ($\epsilon_\text{rec}$) and geometrical acceptance ($A$), needed to correct the raw yield, are determined by applying the reconstruction and selections to simulated events using PYTHIA 8.2 Monash 2013~\cite{PYTHIA_PhysicsAndUsage} in \pp and DPMJET~\cite{DPMJETIII} in \pPb collisions. 
The propagation of particles through the detector is simulated using GEANT 3~\cite{GEANT3UsersGuide}, taking into account the detector conditions during the data taking. 
An additional correction is applied to the cluster energies in the simulation, referred to as MC cluster energy fine tuning, to accurately reproduce the cluster energies observed in data, as described in Ref.~\cite{ALICE:2022qhn}.
The resulting mass and width of the $\omega$ meson extracted in data and simulations are in agreement within the respective statistical uncertainties of each \piz reconstruction method; see supplemental note~\cite{ALICE:SuppNote}.
Furthermore, to correct for a residual mismatch of the material budget in simulation compared to the data, each $\text{V}^0$ candidate in simulation is weighted according to~\cite{ALICE:2023kzv} in eleven radial intervals depending on the conversion point. The weight is then propagated to the $\omega$ meson. The correction shifts the extracted $\omega$ yield by up to 9\% (4.5\%), depending on \pt, and the associated systematic uncertainty is reduced from 9\% (4.5\%) down to 5\% (2.5\%) for the PCM (PCM-EMCal) \piz reconstruction technique.

The geometrical acceptance is calculated as the ratio of the number of accepted $\omega$ mesons (i.e., where all decay particles are within the geometrical limits of the respective detector) to all generated $\omega$ mesons within $\vert y \vert < 0.85$. The reconstruction efficiency is defined as the ratio of reconstructed to accepted $\omega$ mesons in the simulation: $\epsilon_\text{rec} = N^\omega_\text{rec}/N^\omega_\text{acc}$. Here, the number of reconstructed $\omega$ mesons can either be obtained from the background-subtracted signal or by counting \pipipi triplets that are validated to come from the same $\omega$ meson, checked at generator level in the simulation, thereby avoiding the need for background subtraction.
Since both efficiencies were found to agree within their uncertainties and the latter significantly decreases the statistical uncertainties, the \textit{validated} efficiency is utilized in this analysis.
\refFig{Corrections} shows the total correction factor applied to the raw spectra, consisting of a product of the geometrical acceptance, validated efficiency, as well as solid angle normalization, where $\Delta\varphi=2\pi$ and $\Delta y = 1.7$ represent the azimuthal and rapidity ranges in which $\omega$ mesons were produced in the simulations.

While the total correction factors do not differ between the two collision systems, the requirement of either one or both photons converting for the PCM-EMCal and PCM methods reduces their overall reconstruction efficiency by a factor of $1/8.5\%$.
Although this limits the \pt reach of PCM at high \pt, the good energy resolution of PCM photons enables the reconstruction of $\omega$ mesons down to lower \pt.

For the calculation of the nuclear modification factor \RpPb, a correction factor is applied to the \pt spectrum of the $\omega$ measured in pp collisions to account for the boost of the center-of-mass frame in \pPb collisions. 
Using a \Pythia~\cite{PYTHIA_PhysicsAndUsage} simulation, the production is found to be 0.8\% (1.2\%) smaller in the shifted rapidity regime of $-1.315<y<0.385$ compared to the central rapidity range of $\vert y \vert < 0.85$ for low (high) \pt. 
\section{Systematic uncertainties} 
\label{sec:Systematics}

The systematic uncertainties of the $\omega$ meson cross sections and the nuclear modification factor were evaluated as a function of \pt by variations of the selection criteria.
\refTab{Systematics} displays all sources of systematic uncertainties and their magnitude for the three \piz reconstruction methods. 

\begin{sidewaystable}
    \centering
    \begin{tabular}{|l||c:c|c:c|c:c||c:c|c:c|c:c||c:c|c:c|c:c||}
                \hline& \multicolumn{6}{c||}{$\omega$ cross section in \pp} & \multicolumn{6}{c||}{$\omega$ cross section in \pPb} & \multicolumn{6}{c||}{$\omega$ \RpPb}\\\hline
                \piz reconstruction & \multicolumn{2}{c|}{PCM} & \multicolumn{2}{c|}{PCM-EMCal} & \multicolumn{2}{c||}{EMCal} & \multicolumn{2}{c|}{PCM} & \multicolumn{2}{c|}{PCM-EMCal} & \multicolumn{2}{c||}{EMCal} & \multicolumn{2}{c|}{PCM} & \multicolumn{2}{c|}{PCM-EMCal} & \multicolumn{2}{c||}{EMCal}\\\hline
                \pt (\GeVc)               & 3.6 & 9.0 & 3.6 & 9.0 & 3.6 & 9.0 & 3.6 & 9.0 & 3.6 & 9.0 & 3.6 & 9.0 & 3.6 & 9.0 & 3.6 & 9.0 & 3.6 & 9.0 \\\hline\hline
                Signal extraction  & 8.6 & 16.3 & 8.4 & 14.3 & 8.0 & 10.6 & 14.4 & 11.7 & 16.5 & 12.7 & 8.1 & 9.6 & 11.9 & 14.4 & 12.7 & 14.9 & 7.8 & 11.5\\
                Material budget      & 5.0 & 5.0 & 4.6 & 4.6 & 4.2 & 4.2 & 5.0 & 5.0 & 4.6 & 4.6 & 4.2 & 4.2 & - & - & - & - & - & -\\
                Conv. photons & 5.8 & 12.7 & 1.7 & 6.5 & - & - & 9.4 & 8.9 & 1.6 & 4.4 & - & - & 10.5 & 13.3 & 2.6 & 7.6 & - & - \\
                Calo. photons           & - & - & 3.1 & 4.3 & 3.6 & 5.6 & - & - & 3.7 & 4.3 & 6.5 & 3.4 & - & - & 4.4 & 4.7 & 5.4 & 7.5\\
                Neutral pions      & 5.1 & 8.3 & 5.1 & 11.7 & 3.3 & 3.3 & 4.3 & 5.6 & 5.9 & 7.9 & 6.5 & 3.4 & 6.1 & 6.6 & 6.8 & 14.4 & 4.4 & 4.5\\
                Charged pions      & 5.7 & 17.7 & 3.5 & 4.3 & 4.4 & 6.3 & 7.2 & 9.8 & 4.2 & 3.9 & 3.4 & 3.4 & 8.3 & 13.4 & 6.8 & 6.1 & 5.4 & 7.5\\
                Cross section      & 1.8 & 1.8 & 1.8 & 1.8 & 1.8 & 1.8 & 3.5 & 3.5 & 3.5 & 3.5 & 3.5 & 3.5 & 3.9 & 3.9 & 3.9 & 3.9 & 3.9 & 3.9\\
                Branching ratio      & 0.8 & 0.8 & 0.8 & 0.8 & 0.8 & 0.8 & 0.8 & 0.8 & 0.8 & 0.8 & 0.8 & 0.8 & - & - & - & - & - & -\\
                Rapidity shift      & - & - & - & - & - & - & - & - & - & - & - & - & 0.1 & 0.1 & 0.1 & 0.1 & 0.1 & 0.1\\\hline
                Total systematic      & 14.0 & 29.0 & 12.1 & 21.0 & 11.3 & 14.7 & 20.1 & 19.5 & 19.3 & 17.6 & 12.9 & 12.9 & 19.7 & 25.2 & 17.6 & 23.9 & 12.7 & 16.5\\
                Statistical        & 11.8 & 32.2 & 7.6 & 16.5 & 6.1 & 9.6 & 11.8 & 32.2 & 8.5 & 9.9 & 7.1 & 6.7 & 19.5 & 32.2 & 11.5 & 19.5 & 9.5 & 11.9\\\hline
                Combined stat.        & \multicolumn{3}{r:}{4.6} & \multicolumn{3}{l||}{8.1} & \multicolumn{3}{r:}{5.3} & \multicolumn{3}{l||}{5.4} & \multicolumn{3}{r:}{7.1} & \multicolumn{3}{l||}{9.9}\\
                Combined sys.        & \multicolumn{3}{r:}{8.9} & \multicolumn{3}{l||}{13.7} & \multicolumn{3}{r:}{11.3} & \multicolumn{3}{l||}{11.4} & \multicolumn{3}{r:}{11.2} & \multicolumn{3}{l||}{15.3}\\\hline
        \end{tabular}
        \vspace{4mm}
    \caption{Compilation of the relative systematic uncertainties (\%) from the various sources for the \pp and \pPb cross section and the \RpPb using the three \piz reconstruction methods. For \pt-dependent uncertainties, the relative value is given for the two intervals $3.2 < \pt~($GeV$/c)< 4.0$ and $8.0 < \pt~($GeV$/c)< 10.0$, allowing for a comparison between the reconstruction methods in the low- and mid-\pt regime. The systematic uncertainties from different sources are added in quadrature, yielding the total systematic uncertainty. Together with the statistical uncertainties, the cross sections and nuclear modification factors are combined using the BLUE method~\cite{Lista_2017}. The two last rows compile the resulting combined statistical and systematic uncertainties. The two bottom rows show the statistical and systematic uncertainty of the cross sections and the \RpPb after combining the results from the three \piz reconstruction methods for the low- and mid-\pt regime.}
    \label{tab:Systematics}
\end{sidewaystable}

For \pt-dependent sources, the relative uncertainty is given in a low \pt interval ($3.2 < \pt~($GeV$/c)< 4.0$) and an intermediate \pt interval ($8.0 < \pt~($GeV$/c)< 10.0$).
The total systematic uncertainty for each reconstruction method is calculated as the quadratic sum, assuming no correlation between the different uncertainty sources.
Due to the small signal-to-background ratio, the signal extraction represents the dominant contribution of systematic uncertainty for all measured spectra and reconstruction methods, as also found in previous measurements of the $\omega$ meson~\cite{ALICE:2020ylo, ALICE:13TeVomega}.
Six parameters were identified as possible sources of this systematic uncertainty: the choice of the efficiency type, the integration and parametrization ranges, the description of the background and signal, and the choice of the $\lambda$ parameter in the applied $\omega$ meson mass resolution correction, introduced in \refsec{Omega}.
The signal extraction is performed using all possible combinations of reasonable settings for these parameters to estimate this uncertainty accurately. 
The resulting 198 variations of the spectra were found to be normally distributed, allowing for the extraction of the $1\sigma$ uncertainty.

The material budget uncertainty quoted in \refTab{Systematics} accounts for the accuracy of the detector materials description in the simulation. 
The material uncertainties per EMCal photon and conversion photon are 2.1\% and 2.5\%, respectively~\cite{ALICE:2017ryd, ALICE:2023kzv}.

Uncertainties stemming from the reconstruction of conversion and calorimeter photons (see conv. and calo. photons in \reftab{Systematics}), as well as from neutral and charged pions, were estimated as described in Refs.~\cite{ALICE:13TeVpi0, ALICE:13TeVomega} through variations of the selection criteria introduced in \refsec{Pions}.

For \RpPb, correlated uncertainties between the two collision systems cancel out by simultaneously varying the selection criteria and signal extraction in the two contributing cross sections.

The uncertainties of the visible \ac{MB} cross section in \pp and \pPb of 1.8\%~\cite{ALICE:2018lum} and 3.5\%~\cite{ALICE:2014gvw}, respectively, were taken from their corresponding publications. For the uncertainty in the branching ratio $\omega\rightarrow\pip\pim\gamma\gamma$, the uncertainties of the two consecutive decays of the $\omega$ and \piz meson quoted by the \ac{PDG}~\cite{PDG} are added in quadrature.
The rapidity shift applied to the \pp reference measurement for the \RpPb is assumed to introduce a constant uncertainty of 0.1\%, which corresponds to 10\% of the shift, taking into consideration the statistical uncertainty and introduced \ac{MC} dependence.

The systematic uncertainties are found to be fully correlated between \pt intervals for all systematic uncertainties except the signal extraction uncertainty, where the correlation between \pt intervals is assumed to be negligible. 

The last row of \refTab{Systematics} shows the systematic uncertainty of the combined cross section or nuclear modification factor. These combined values are obtained using the Best Linear Unbiased Estimate (BLUE) method~\cite{Lista_2017}, which takes into account correlations between the statistical and systematic uncertainties of the individual reconstruction methods. The correlation coefficients were evaluated for each source of uncertainty. For example, the material budget uncertainty between PCM and EMCal is fully uncorrelated, while the systematic uncertainty of the branching ratio is fully correlated.
\section{Results} 
\label{sec:Results}
\subsection{Production cross section of \texorpdfstring{$\omega$}{omega} mesons}

The \pt differential Lorentz-invariant production cross section of $\omega$ mesons in \pp and \pPb collisions is calculated for all three \piz reconstruction methods as 
\begin{equation}
    E \frac{\text{d}^3 \sigma^\omega}{\text{d}^3 p} = \frac{1}{\mathcal{L}_\text{int}} \frac{1}{2 \pi \Delta y A \epsilon_\text{rec}} \frac{1}{\mathcal{B}} \frac{N^\omega}{\pt\Delta \pt},
\end{equation}
where $\mathcal{L}_\text{int}$ is the integrated luminosity in the respective collision system introduced in \refsec{Datasets}. The number of reconstructed $\omega$ mesons $N^\omega$ is furthermore normalized by the total correction factor $\epsilon~=~2\pi\Delta y A \epsilon_\text{rec}$ discussed in \refsec{Omega}, its \pt value and the width of the \pt interval $\Delta$\pt, as well as the branching ratio $\mathcal{B}(\omega\rightarrow \pip\pim\gamma\gamma) = (88.15\pm0.70)\%$.

\begin{figure}[tb]
    \centering
    \includegraphics[width = 0.495\textwidth]{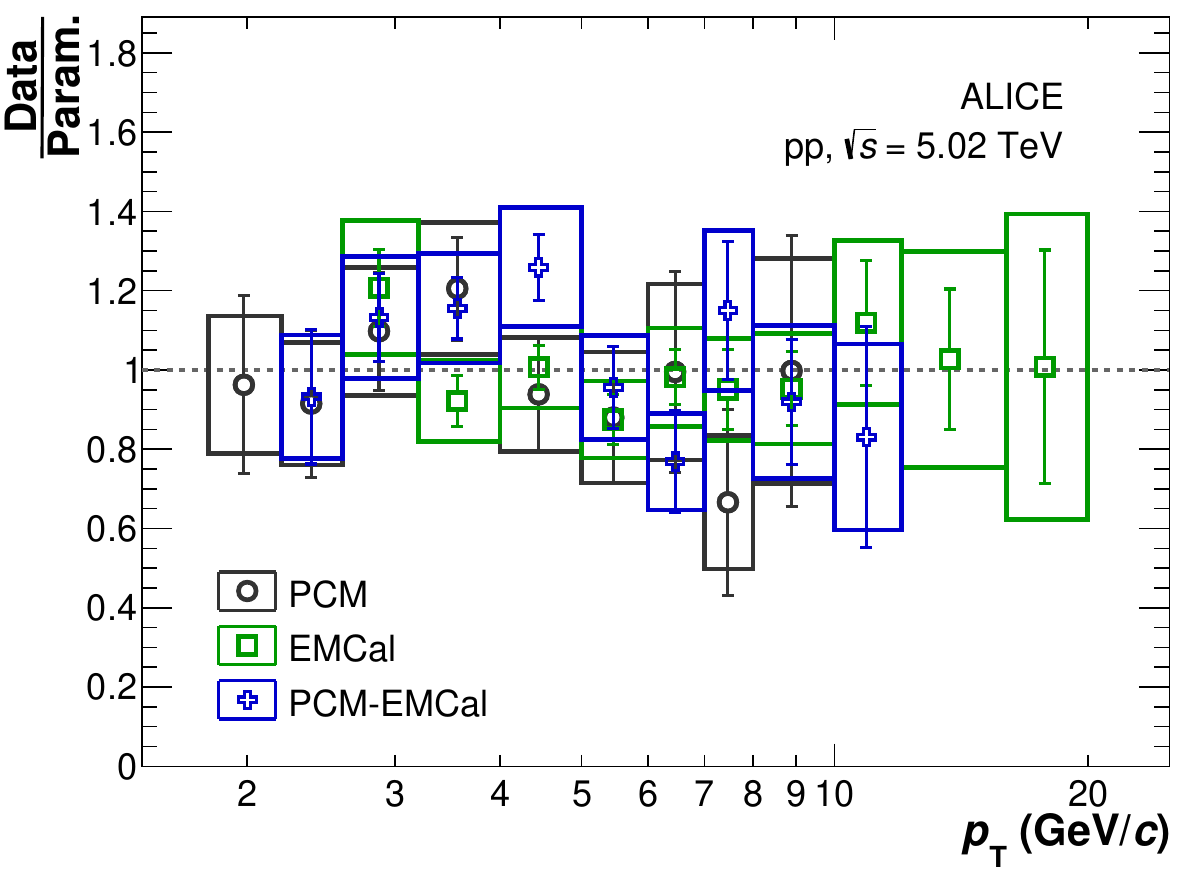}
    \includegraphics[width = 0.495\textwidth]{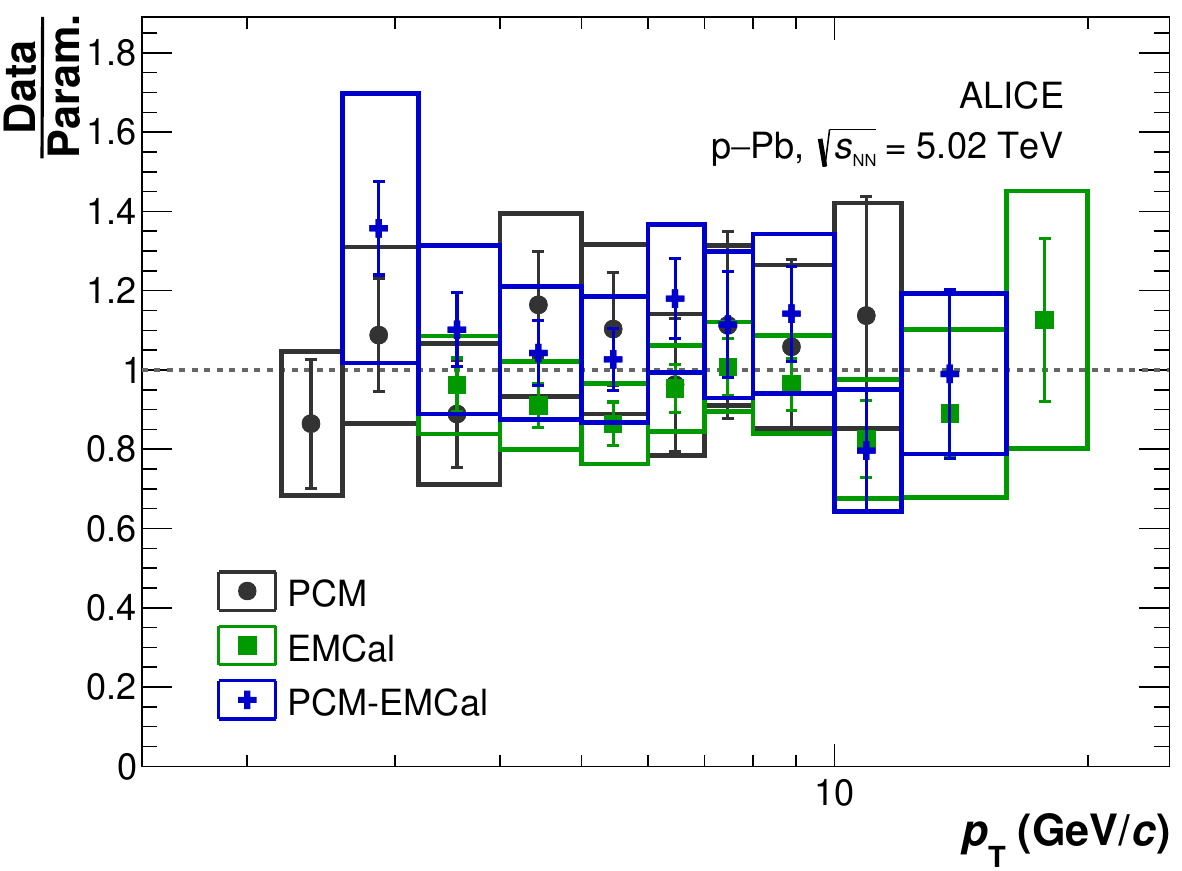}
    \caption{Ratio of the $\omega$ meson production cross sections extracted using the three different \piz reconstruction methods to a combined \LT parametrization. Both in \pp (left) and in \pPb (right), agreement between the reconstruction methods is observed within the statistical (bars) and systematic (boxes) uncertainties.}
    \label{fig:RatiosToTsallis}
\end{figure}

\refFig{RatiosToTsallis} shows the ratio of the $\omega$ meson cross sections extracted using the three \piz reconstruction methods to a combined \LT parametrization, as defined in Ref.~\cite{ALICE:13TeVomega}. Agreement is observed between the extracted cross sections within their respective statistical and systematic uncertainties. While the overlap of different \piz reconstruction methods reduces the statistical and systematic uncertainties, \reffig{RatiosToTsallis} highlights their complementarity, extending the \pt coverage of the measurement. 
The three production cross sections, and the respective \RpPb's, for each collision system are combined using the BLUE method~\cite{Lista_2017}, taking into consideration the correlation between the uncertainties, as done in previous measurements~\cite{ALICE:2020ylo, ALICE:13TeVomega}.

\begin{figure}[tb]
    \centering
    \includegraphics[width = 0.5\textwidth]{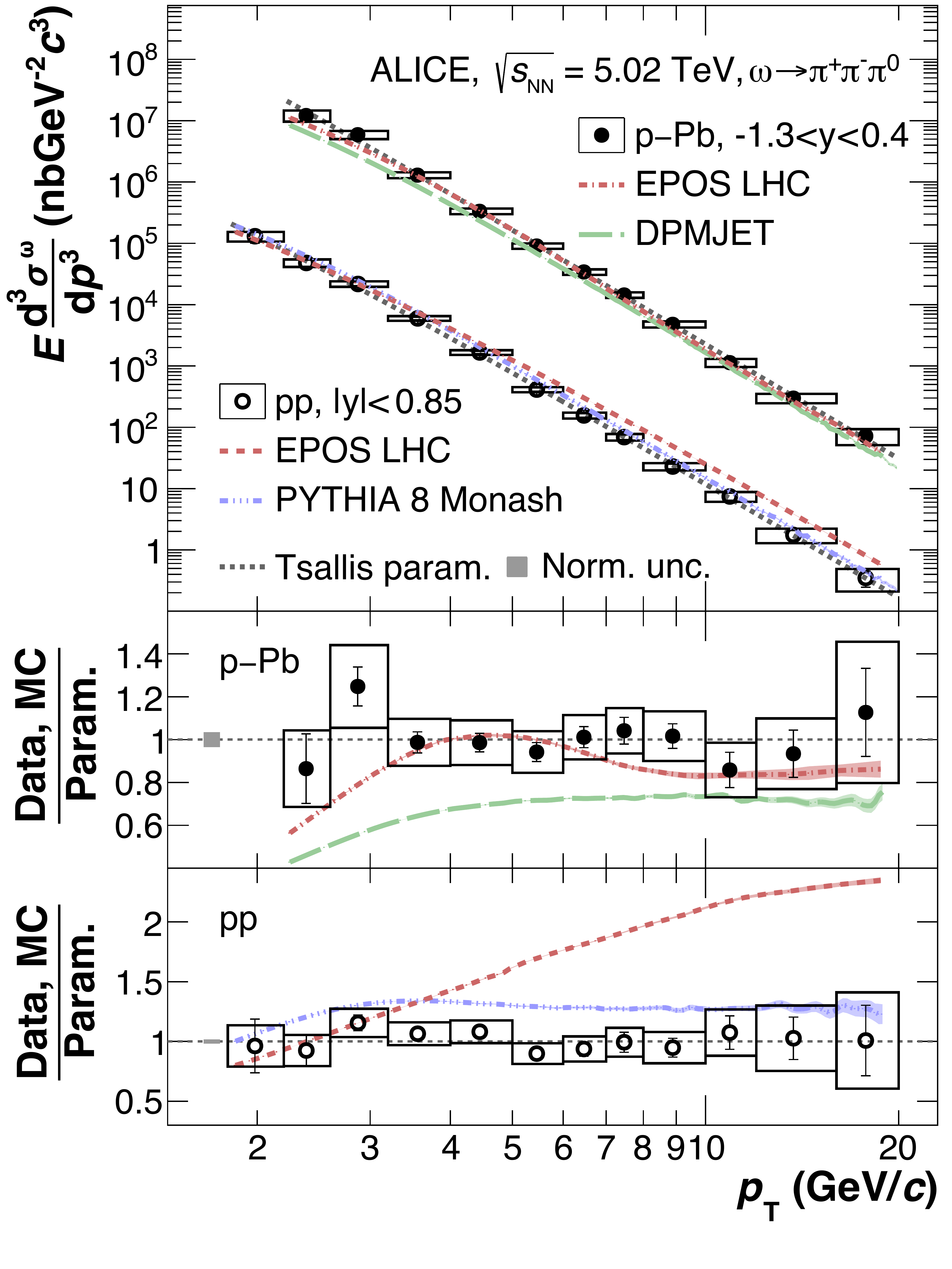}
    \caption{Lorentz-invariant $\omega$ meson production cross section in \pp (open markers) and \pPb (closed markers). Statistical uncertainties are represented by vertical error bars, while boxes show the systematic uncertainties. Furthermore shown are \LT parametrizations of the two cross sections and two predictions from simulations of the $\omega$ meson production per collision system. The lower panels contain the ratios of the data and simulations to the parametrization of the data points in the respective collision system. A gray box in each ratio panel depicts the normalization uncertainty of the visible minimum bias cross section.}
    \label{fig:XSecs}
\end{figure}

The combined \pt differential Lorentz-invariant production cross section of $\omega$ mesons is extracted in \pp and \pPb collisions at \fivenn within the transverse momentum range of 1.8~$\leq$~\pt~(\GeVc)~$<$~20 and 2.2~$\leq$~\pt(\GeVc)~$<$~20, respectively. The \pp and \pPb measurements are shown in \reffig{XSecs}, covering the rapidity interval of $\vert y \vert$~$<$~0.85 and -1.315~$<$~y~$<$~0.385, respectively. 
To account for the finite width of the \pt-intervals, the horizontal positions of the data points shown in \reffig{XSecs} are shifted towards lower \pt, according to the method outlined in Ref.~\cite{LAFFERTY1995541}. The description of the data points using a \LT parametrization, analogous to Ref.~\cite{ALICE:13TeVomega}, yielded \pt-shifts on the order of 1\% towards lower \pt. 
This bin-shift is performed in y-coordinates for the $\omega/\pi^0$ ratios and \RpPb where, due to the similar shape of the spectra, the shift is below 1\% for all \pt.

\refFig{XSecs} also contains four predictions from simulations using different MC generators and a \LT parametrization of the data for each collision system. 
The $\chi^2$/ndf of the parametrization is 0.49 (0.48) in \pp (\pPb), taking into account the total uncertainties.
The lower panels depict the ratio of data and the predictions from the event generators to the \LT parametrization.
The DPMJET simulation~\cite{DPMJETIII}, based on the \ac{DPM}, 
roughly describes the \pt dependence of the cross section in \pPb, but it underestimates the production of $\omega$ mesons by approximately 30\%. 
On the other hand, the production of $\omega$ mesons in \pPb collisions is well described by EPOS LHC~\cite{EPOSLHC}. The EPOS model describes hadronic interactions following a quantum mechanical multiple-scattering approach based on partons and strings with built-in collective hadronization~\cite{EPOS}. The EPOS LHC adaptation~\cite{EPOSLHC} includes modifications, aiming to describe the first LHC results. 
However, when applied to \pp collisions, the EPOS LHC simulations do not describe the data, with discrepancies of up to 100\%, as shown in the lower panel.
While the other \pp simulation, PYTHIA 8.2 with the Monash 2013 tune~\cite{PYTHIA_PhysicsAndUsage}, describes the shape of the production cross section in \pp collisions, it overestimates the production of $\omega$ mesons by about 30\%.
A similar deviation between these \Pythia predictions and the respective measured cross section was observed in previous $\omega$ meson analyses at different collision energies~\cite{ALICE:2020ylo, ALICE:13TeVomega}, hinting at a possible energy-independent overestimation of the $\omega$ meson production in the Monash 2013 tune of \Pythia.

\subsection{\texorpdfstring{$\omega/\pi^0$}{omega/pi0} ratios}

\begin{figure}
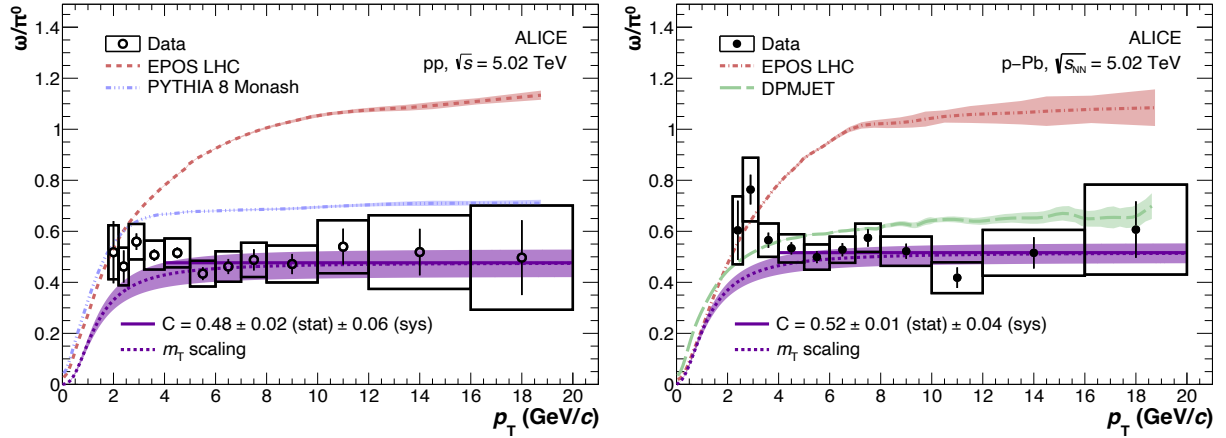

        \includegraphics[width=0.495\textwidth]{figures_paper/OmegaToPiZero_mT_pp.pdf}
    \hfill
        \includegraphics[width=0.495\textwidth]{figures_paper/OmegaToPiZero_mT_pPb.pdf}
    \caption{Production ratio $\omega$/\piz in \pp and \pPb collisions calculated using the $\omega$ meson production measurements and the corresponding charged~\cite{ALICE:2019hno} and neutral pion~\cite{ALICE:2018vhm} references, respectively. The charged pion reference was scaled up by 3.3\% (see text for details). Also shown are two predictions from simulations per collision system, with bands representing their statistical uncertainties, as well as a $m_\text{T}$ scaling curve, which converges towards a high-\pt constant (C) fitted to the ratio for \pt~$>$~4~\GeVc.}
    \label{fig:MyOTPs}
\end{figure}

In \pPb collisions, the ratio of produced $\omega$ and \piz mesons is calculated from the measured $\omega$ meson cross section and the corresponding \piz spectrum at \fivenn~\cite{ALICE:2018vhm}. 
To obtain a \piz reference for the \pp measurement, the published spectrum of charged pions~\cite{ALICE:2019hno} is used as a proxy. To account for the slightly higher production yield of neutral pions mainly due to the isospin symmetry breaking $\eta\rightarrow\piz\piz\piz$ decay, the measured charged pion spectrum is scaled up by 3.3\%. This surplus is determined from the \mbox{(\pip+\pim)/(2\piz)} ratio in \pp collisions at \fivenn simulated using \Pythia~\cite{PYTHIA_PhysicsAndUsage} via a constant parametrization for 1.8~$\leq$~\pt~(\GeVc)~$<$~20. An additional relative systematic uncertainty of 3.3\% is added to the $\omega$/$\pi^0$ ratio in \pp collisions to account for the model dependence introduced. Correlated systematic uncertainties of the visible \ac{MB} cross section, and in the case of \pp the material budget, are hereby removed. All other systematic and statistical uncertainties are assumed to be uncorrelated. 
\refFig{MyOTPs} shows the measured $\omega$/\piz ratio in the two collision systems; neither of them display a significant \pt dependence. 

Taking into consideration the bin-by-bin correlation between the systematic uncertainties, an average $\omega/\piz$ constant for \pt~$>$~4~\GeVc is evaluated following the procedure described in Ref.~\cite{Lista_2017}, yielding $C^{\omega/\pi^0}_\text{pp\phantom{b}} =~0.48~\pm~0.02~\text{(stat.)}~\pm~0.06~\text{(sys.)}$ in \pp collisions, and $C^{\omega/\pi^0}_\text{pPb} =~0.52~\pm~0.01~\text{(stat.)}~\pm~0.04~\text{(sys.)}$ in \pPb collisions.
Within the given uncertainties and \pt reach, the $\omega$/\piz production ratios agree with one another, suggesting the production ratio to be independent of CNM effects within the given uncertainties and inspected \pt range.

The extracted high-\pt constants can furthermore be used as input for the transverse mass scaling prediction of the $\omega$/\piz ratio. This empirical scaling relation describes the production ratio by assuming that the production of all mesons follows the same fundamental underlying function of the particle's transverse mass ($m_\text{T} = \sqrt{p_\text{T}^2+m_\text{inv}^2}$), scaled by a constant parameter $C$~\cite{Altenk_mper_2017}. From the extracted high-\pt constants, this assumption is used to derive the \pt dependence of the particle ratio, as shown as purple bands in \reffig{MyOTPs}. In both collision systems, a slight tension between the $\omega$/\piz ratio prediction assuming $m_\text{T}$ scaling and the data is observed at low transverse momenta \pt$\approx3~$\GeVc, where $m_\text{T}$ scaling suggests a decrease of the $\omega$/\piz ratio, which is not observed in the data within uncertainties. A slight discrepancy is expected, as $m_\text{T}$ scaling was found to be broken for LHC energies due to the feed-down of decays into the \piz~\cite{Altenk_mper_2017}. However, the increased number of neutral pions from the feed-down of higher mass particles would decrease the $\omega$/\piz ratio compared to the $m_\text{T}$ scaling prediction. The excess of the $\omega$/\piz ratio at low \pt compared to the $m_\text{T}$ scaling was also observed at \thirteen, but its underlying cause remains unexplained. This discrepancy shows the need for $\omega$ meson production measurements such as this one, as studies of direct photons or dileptons commonly resort back to this $m_\text{T}$ scaling prediction to evaluate hadronic feed-down contributions due to a lack of experimental data~\cite{ALICE:2015xmh, ALICE:2018gev}.

In addition to the measured production ratios, \reffig{MyOTPs} also includes predictions of the $\omega$/\piz ratio by three different event generators. The predictions of the $\omega$/\piz ratio by EPOS LHC~\cite{EPOSLHC}, shown in red, are very similar between \pp and \pPb, suggesting the system size does not have a strong impact on the relative hadronization fractions into $\omega$ mesons and neutral pions. However, the production ratio is overestimated by about 100\% in both collision systems. So, while EPOS LHC describes the production of $\omega$ mesons in p--Pb and not in pp collisions, see \reffig{XSecs}, the \piz production is described only in pp and not in \pPb collisions. This inadequate collision system and particle species dependence of the EPOS LHC predictions possibly hints at a lack of experimental data for tuning the collective hadronization implemented in EPOS LHC~\cite{EPOS}. 
The similar production ratio predictions by \Pythia~\cite{PYTHIA_PhysicsAndUsage} in \pp and DPMJET~\cite{DPMJETIII} in \pPb can be explained by the fact that the hadronization implemented in DPMJET is based on the Lund string model, also used by the \Pythia event generator. 
Both generators using the Lund string model overestimate the $\omega$/\piz ratio by about $10-20\%$.
It is, however, not evident whether the different hadronization descriptions cause this more accurate description of the $\omega$/\piz ratio compared to EPOS LHC, or whether this difference should be attributed to the different tunes and experimental input used.

\begin{figure}
        \includegraphics[width=0.633\textwidth]{figures_paper/Full_ppOmegaToPiZero.pdf}
    \hfill
        \includegraphics[width=0.36\textwidth]{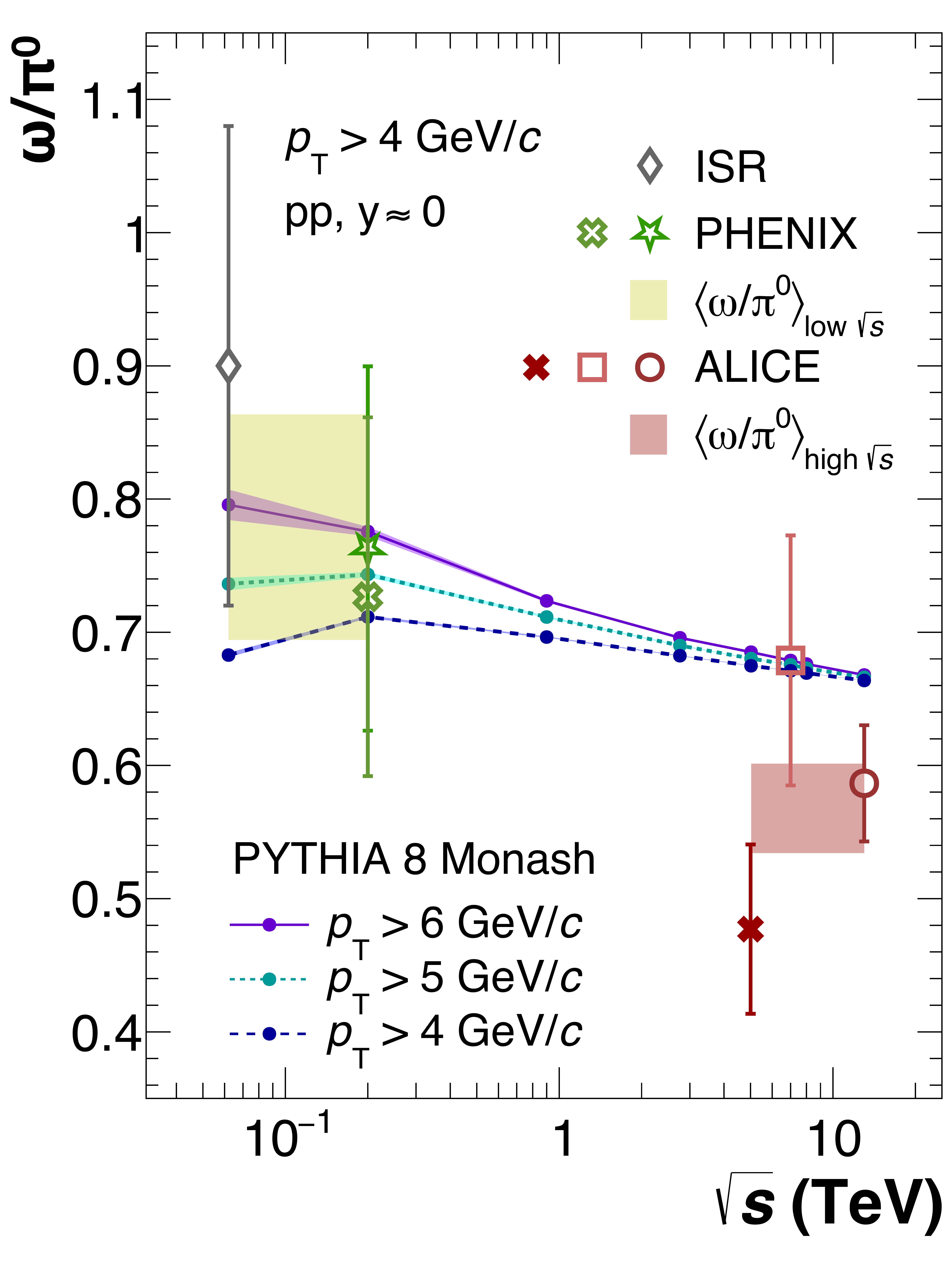}
    \caption{Compilation of measurements of the $\omega$/\piz ratio (left) in \pp collisions at various center-of-mass energies covering 0.062~$<$~\s~(\TeV)~$<$~13 at the ISR~\cite{ISR_omega_pp}, PHENIX~\cite{PHENIX_omega_pp} and \acs{ALICE}~\cite{ALICE:2020ylo,ALICE:13TeVomega}, including the \pp measurement presented herein. Vertical error bars represent total uncertainties, high-\pt data points at \thirteen up to \pt~=~50~\GeVc are omitted here for better visibility. The right-hand side displays values obtained from parametrizations of the $\omega/\piz$ ratios above \pt~$>$~4~\GeVc, as well as predictions of this high-\pt constant for different lower bounds using \Pythia~\cite{PYTHIA_PhysicsAndUsage}.}
    \label{fig:AllOTPs}
\end{figure}

\refFig{AllOTPs} shows the measured $\omega$/\piz ratio in pp collisions, compared to previous measurements, covering almost three orders of magnitude in center-of-mass energies.
For the data points of the ISR measurements, only a minimum \pt was defined in the measurement. The \pt positions of these data points in a given \pt interval were therefore set to the expectation value of the \LT parametrization of the $\omega$ cross section in \pp collisions at $\sqrt{s}=$~62~\GeV, taken from \Pythia~\cite{PYTHIA_PhysicsAndUsage}.

The $\omega$/\piz ratios measured at different energies are compatible within the given total uncertainties represented by vertical bars. A slight tension is visible when considering the entire \pt range of the measurement at \twoHnn, depicted with green markers, and at \fivenn, shown with red markers. To better visualize this possible difference of the $\omega$/\piz ratio for different center-of-mass energies, the ratios shown in \reffig{AllOTPs} are parametrized with a constant for \pt~$>4$~\GeVc, with the resulting values of these high-\pt fits of the $\omega$/\piz ratio compiled in \reffig{AllOTPs}, right panel. As the bin-by-bin correlation of the systematic uncertainty for the cited measurements is unknown, their systematic uncertainties are assumed to be fully correlated between the bins for calculating the uncertainty of the high-\pt constant. This approach provides an upper-limit estimation of the total uncertainties shown in \reffig{AllOTPs}. To isolate possible effects of the center-of-mass energy, only $\omega$/\piz ratios measured in \acs{pp} collisions are considered for this comparison. 
The resulting high-\pt $\omega$/\piz constant as a function of the center-of-mass energy presented in \reffig{AllOTPs} displays a slight tendency towards a lower $\omega$/\piz ratio with rising center-of-mass energy.
The significance of this effect is estimated by calculating an averaged ratio at low center-of-mass energy (62~$<$~\s~(\GeV)~$<$~200) and one at high center-of-mass energy (5.02~$<$~\s~(\TeV)~$<$~13). By assuming all uncertainties between the different analyses to be fully uncorrelated, the averaged ratios are $\langle \omega/\piz \rangle_\text{low $\sqrt{s}$} = 0.78\pm0.09$ and $\langle \omega/\piz \rangle_\text{high $\sqrt{s}$} = 0.56\pm0.04$, which are shown as colored bands in \reffig{AllOTPs}. This corresponds to a difference of the ratios at low and high center-of-mass energies with a significance of 2.2~$\sigma$.

The trend of a decreasing $\omega/\piz$ ratio with rising center-of-mass energy is also observed in \Pythia simulations~\cite{PYTHIA_PhysicsAndUsage}, shown in blue, cyan, and purple lines for different \pt-thresholds in \reffig{AllOTPs}. In the simulations, the production of heavier particles at larger center-of-mass energies is found to result in more hadronic feed-down into primary neutral pions, thereby decreasing the $\omega/\piz$ ratio.
Further precision measurements at both low and high collision energies are needed to investigate this trend.

\subsection{Nuclear modification factor \texorpdfstring{$R_\text{pPb}$}{RpPb}}

\begin{figure}[tb]
    \centering
    \includegraphics[width = 0.633\textwidth]{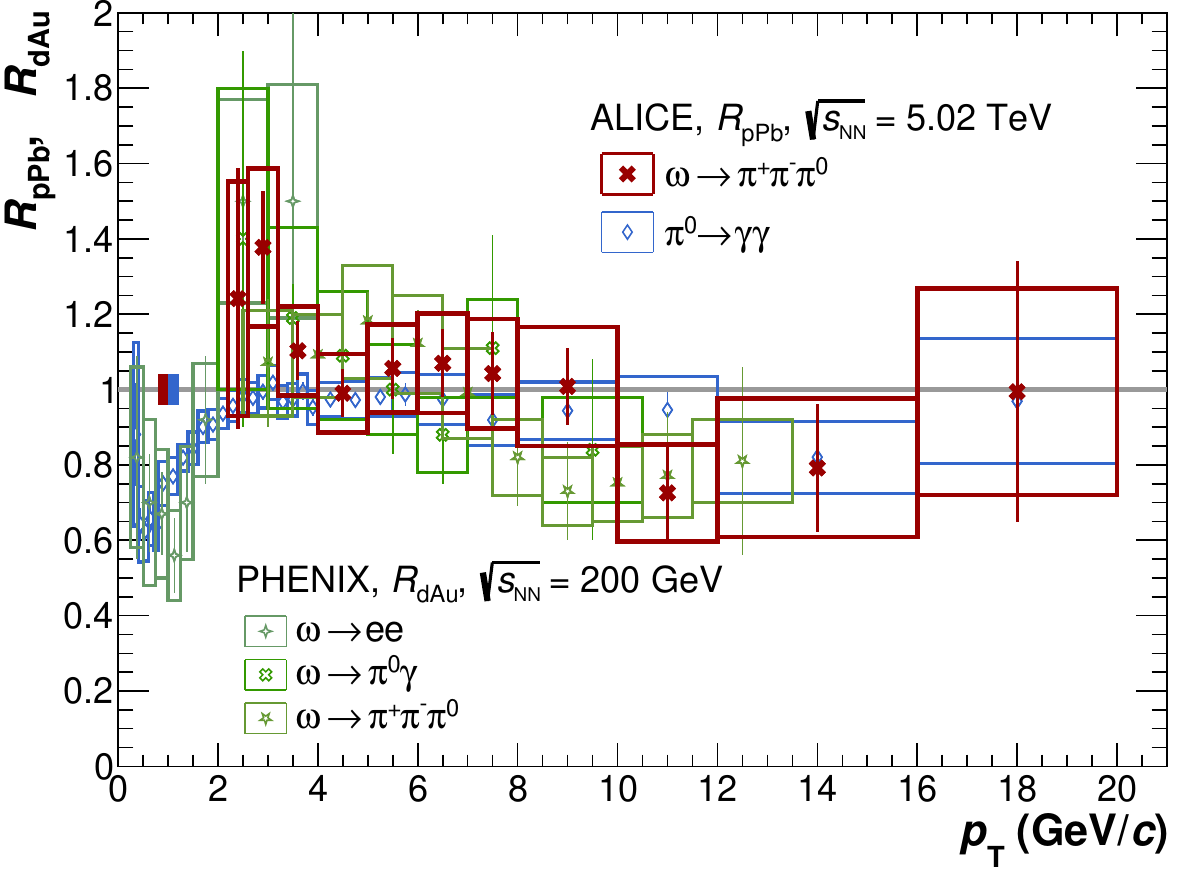}
    \caption{Measured nuclear modification factor \RpPb of the $\omega$ meson at \fivenn shown with red markers, as well as a nuclear modification factor of neutral pions at the same center-of-mass energy~\cite{ALICE:2018vhm} and the measurements for $\omega$ mesons at \twoHnn~\cite{PHENIX_omega_full}. Vertical error bars and boxes represent statistical and systematic uncertainties, while the two solid boxes show the normalization uncertainty of the ALICE measurements.}
    \label{fig:RpA}
\end{figure}

The nuclear modification factor of the $\omega$ meson production at \fivenn, extracted in the transverse momentum range of 2.2~$\leq$~\pt(\GeVc)~$<$~20 and the rapidity interval of -1.315~$<$~y~$<$~0.385, is shown with red markers in \reffig{RpA}. 
This represents the first measurement of the nuclear modification factor of the $\omega$ meson at LHC energies. Within the statistical and systematic uncertainties, represented by vertical bars and boxes, respectively, the nuclear modification factor is found to be compatible with unity over the measured \pt range. Consequently, no CNM effects on the $\omega$ meson production are observed within the given uncertainties.

\refFig{RpA} also displays the nuclear modification factor for neutral pions, measured at the same center-of-mass energy of \fivenn~\cite{ALICE:2018vhm}. The visible agreement between the nuclear modification factors for the light \piz and the six times heavier $\omega$ meson~\cite{PDG} implies that, within the \pt range and the given uncertainties of the two measurements, no mass dependence of the nuclear modification factor is observed.

Finally, \reffig{RpA} also includes the nuclear modification factors \RdAu for the production of $\omega$ mesons at \twoHnn~\cite{PHENIX_omega_full} in d--Au collisions, measured in different decay channels. These nuclear modification factors \RdAu are compatible with the herein presented \RpPb within the respective statistical and systematic uncertainties. 

\section{Conclusion} 
\label{sec:Conclusion}
The \pt-differential invariant cross sections of $\omega$ mesons in \pp and \pPb collisions at $\sqrt{s_{\text{NN}}}$~=~5.02~TeV have been presented. The measurements, performed using the $\omega \rightarrow \pi^+\pi^-\pi^0$ decay channel, cover a transverse momentum range of 1.8~$\leq$~\pt~(\GeVc)~$<$~20 and 2.2~$\leq$~\pt~(\GeVc)~$<$~20, respectively. Three partially independent reconstruction methods were used for the reconstruction of the neutral pion, significantly reducing the systematic uncertainties of the combined $\omega$ cross section.
Predictions from \ac{MC} event generators in both \pp and \pPb collisions do not describe the $\omega$ production cross section within the given uncertainties. While EPOS LHC describes the spectrum in \pPb reasonably well, it overestimates the data by up to 100\% in \pp collisions. \Pythia overestimates the spectrum in pp as previously seen in \pp at $\sqrt{s}$~=~13~TeV, and predictions from DPMJET in \pPb collisions underestimate the spectrum.
The $\omega$/$\pi^0$ ratio agrees between \pp and \pPb collisions, indicating no system size dependence of the ratio, as also observed for the $\eta/\piz$ ratio~\cite{ALICE:2018vhm}. Comparisons between measurements of the $\omega$/$\pi^0$ in \pp collisions for energies up to $\sqrt{s}$~=~200~GeV and LHC energies suggest an energy dependence of the ratio for \pt~$>$~4~\GeVc with a significance of 2.2~$\sigma$. In \Pythia, showing qualitatively the same dependence, this trend is attributed to a rising feed-down contribution into the $\pi^0$ spectrum with rising collision energy. Finally, the first \RpPb for $\omega$ mesons at LHC energies was presented for 2.2~$\leq$~\pt~(\GeVc)~$<$~20. Agreeing with unity over the full \pt range, the data does not show modification due to cold nuclear matter. Furthermore, the data agrees with measurements of $\pi^0$ at the same collision energy, as well as with a measurement of $\omega$ mesons in d--Au collisions at $\sqrt{s_{\text{NN}}}$~=~200~GeV. 


\newenvironment{acknowledgement}{\relax}{\relax}
\begin{acknowledgement}
\section*{Acknowledgements}

The ALICE Collaboration would like to thank all its engineers and technicians for their invaluable contributions to the construction of the experiment and the CERN accelerator teams for the outstanding performance of the LHC complex.
The ALICE Collaboration gratefully acknowledges the resources and support provided by all Grid centres and the Worldwide LHC Computing Grid (WLCG) collaboration.
The ALICE Collaboration acknowledges the following funding agencies for their support in building and running the ALICE detector:
A. I. Alikhanyan National Science Laboratory (Yerevan Physics Institute) Foundation (ANSL), State Committee of Science and World Federation of Scientists (WFS), Armenia;
Austrian Academy of Sciences, Austrian Science Fund (FWF): [M 2467-N36] and Nationalstiftung f\"{u}r Forschung, Technologie und Entwicklung, Austria;
Ministry of Communications and High Technologies, National Nuclear Research Center, Azerbaijan;
Conselho Nacional de Desenvolvimento Cient\'{\i}fico e Tecnol\'{o}gico (CNPq), Financiadora de Estudos e Projetos (Finep), Funda\c{c}\~{a}o de Amparo \`{a} Pesquisa do Estado de S\~{a}o Paulo (FAPESP) and Universidade Federal do Rio Grande do Sul (UFRGS), Brazil;
Bulgarian Ministry of Education and Science, within the National Roadmap for Research Infrastructures 2020-2027 (object CERN), Bulgaria;
Ministry of Education of China (MOEC) , Ministry of Science \& Technology of China (MSTC) and National Natural Science Foundation of China (NSFC), China;
Ministry of Science and Education and Croatian Science Foundation, Croatia;
Centro de Aplicaciones Tecnol\'{o}gicas y Desarrollo Nuclear (CEADEN), Cubaenerg\'{\i}a, Cuba;
Ministry of Education, Youth and Sports of the Czech Republic, Czech Republic;
The Danish Council for Independent Research | Natural Sciences, the VILLUM FONDEN and Danish National Research Foundation (DNRF), Denmark;
Helsinki Institute of Physics (HIP), Finland;
Commissariat \`{a} l'Energie Atomique (CEA) and Institut National de Physique Nucl\'{e}aire et de Physique des Particules (IN2P3) and Centre National de la Recherche Scientifique (CNRS), France;
Bundesministerium f\"{u}r Bildung und Forschung (BMBF) and GSI Helmholtzzentrum f\"{u}r Schwerionenforschung GmbH, Germany;
General Secretariat for Research and Technology, Ministry of Education, Research and Religions, Greece;
National Research, Development and Innovation Office, Hungary;
Department of Atomic Energy Government of India (DAE), Department of Science and Technology, Government of India (DST), University Grants Commission, Government of India (UGC) and Council of Scientific and Industrial Research (CSIR), India;
National Research and Innovation Agency - BRIN, Indonesia;
Istituto Nazionale di Fisica Nucleare (INFN), Italy;
Japanese Ministry of Education, Culture, Sports, Science and Technology (MEXT) and Japan Society for the Promotion of Science (JSPS) KAKENHI, Japan;
Consejo Nacional de Ciencia (CONACYT) y Tecnolog\'{i}a, through Fondo de Cooperaci\'{o}n Internacional en Ciencia y Tecnolog\'{i}a (FONCICYT) and Direcci\'{o}n General de Asuntos del Personal Academico (DGAPA), Mexico;
Nederlandse Organisatie voor Wetenschappelijk Onderzoek (NWO), Netherlands;
The Research Council of Norway, Norway;
Pontificia Universidad Cat\'{o}lica del Per\'{u}, Peru;
Ministry of Science and Higher Education, National Science Centre and WUT ID-UB, Poland;
Korea Institute of Science and Technology Information and National Research Foundation of Korea (NRF), Republic of Korea;
Ministry of Education and Scientific Research, Institute of Atomic Physics, Ministry of Research and Innovation and Institute of Atomic Physics and Universitatea Nationala de Stiinta si Tehnologie Politehnica Bucuresti, Romania;
Ministry of Education, Science, Research and Sport of the Slovak Republic, Slovakia;
National Research Foundation of South Africa, South Africa;
Swedish Research Council (VR) and Knut \& Alice Wallenberg Foundation (KAW), Sweden;
European Organization for Nuclear Research, Switzerland;
Suranaree University of Technology (SUT), National Science and Technology Development Agency (NSTDA) and National Science, Research and Innovation Fund (NSRF via PMU-B B05F650021), Thailand;
Turkish Energy, Nuclear and Mineral Research Agency (TENMAK), Turkey;
National Academy of  Sciences of Ukraine, Ukraine;
Science and Technology Facilities Council (STFC), United Kingdom;
National Science Foundation of the United States of America (NSF) and United States Department of Energy, Office of Nuclear Physics (DOE NP), United States of America.
In addition, individual groups or members have received support from:
Czech Science Foundation (grant no. 23-07499S), Czech Republic;
FORTE project, reg.\ no.\ CZ.02.01.01/00/22\_008/0004632, Czech Republic, co-funded by the European Union, Czech Republic;
European Research Council (grant no. 950692), European Union;
Deutsche Forschungs Gemeinschaft (DFG, German Research Foundation) ``Neutrinos and Dark Matter in Astro- and Particle Physics'' (grant no. SFB 1258), Germany;
ICSC - National Research Center for High Performance Computing, Big Data and Quantum Computing and FAIR - Future Artificial Intelligence Research, funded by the NextGenerationEU program (Italy).
\end{acknowledgement}

\bibliographystyle{utphys}   
\bibliography{bibliography, AlicePaper}

\newpage
\appendix

%
%

\section{The ALICE Collaboration}
\label{app:collab}
\begin{flushleft} 
\small

S.~Acharya\,\orcidlink{0000-0002-9213-5329}\,$^{\rm 50}$, 
A.~Agarwal$^{\rm 133}$, 
G.~Aglieri Rinella\,\orcidlink{0000-0002-9611-3696}\,$^{\rm 32}$, 
L.~Aglietta\,\orcidlink{0009-0003-0763-6802}\,$^{\rm 24}$, 
M.~Agnello\,\orcidlink{0000-0002-0760-5075}\,$^{\rm 29}$, 
N.~Agrawal\,\orcidlink{0000-0003-0348-9836}\,$^{\rm 25}$, 
Z.~Ahammed\,\orcidlink{0000-0001-5241-7412}\,$^{\rm 133}$, 
S.~Ahmad\,\orcidlink{0000-0003-0497-5705}\,$^{\rm 15}$, 
S.U.~Ahn\,\orcidlink{0000-0001-8847-489X}\,$^{\rm 71}$, 
I.~Ahuja\,\orcidlink{0000-0002-4417-1392}\,$^{\rm 36}$
% ...

\section*{Affiliation Notes}

$^{\rm I}$ Deceased\\
$^{\rm II}$ Also at: Max-Planck-Institut fur Physik, Munich, Germany\\
$^{\rm III}$ Also at: Italian National Agency for New Technologies, Energy and Sustainable Economic Development (ENEA), Bologna, Italy\\
$^{\rm IV}$ Also at: Dipartimento DET del Politecnico di Torino, Turin, Italy\\
$^{\rm V}$ Also at: Department of Applied Physics, Aligarh Muslim University, Aligarh, India\\
$^{\rm VI}$ Also at: Institute of Theoretical Physics, University of Wroclaw, Poland\\
$^{\rm VII}$ Also at: Facultad de Ciencias, Universidad Nacional Aut\'{o}noma de M\'{e}xico, Mexico City, Mexico\\

\section*{Collaboration Institutes}

$^{1}$ A.I. Alikhanyan National Science Laboratory (Yerevan Physics Institute) Foundation, Yerevan, Armenia\\
$^{2}$ AGH University of Krakow, Cracow, Poland\\
$^{3}$ Bogolyubov Institute for Theoretical Physics, National Academy of Sciences of Ukraine, Kiev, Ukraine\\
$^{4}$ Bose Institute, Department of Physics  and Centre for Astroparticle Physics and Space Science (CAPSS), Kolkata, India\\
$^{5}$ California Polytechnic State University, San Luis Obispo, California, United States\\
$^{6}$ Central China Normal University, Wuhan, China\\
$^{7}$ Centro de Aplicaciones Tecnol\'{o}gicas y Desarrollo Nuclear (CEADEN), Havana, Cuba\\
$^{8}$ Centro de Investigaci\'{o}n y de Estudios Avanzados (CINVESTAV), Mexico City and M\'{e}rida, Mexico\\
$^{9}$ Chicago State University, Chicago, Illinois, United States\\
$^{10}$ China Nuclear Data Center, China Institute of Atomic Energy, Beijing, China\\
$^{11}$ China University of Geosciences, Wuhan, China\\
$^{12}$ Chungbuk National University, Cheongju, Republic of Korea\\
$^{13}$ Comenius University Bratislava, Faculty of Mathematics, Physics and Informatics, Bratislava, Slovak Republic\\
$^{14}$ Creighton University, Omaha, Nebraska, United States\\
$^{15}$ Department of Physics, Aligarh Muslim University, Aligarh, India\\
$^{16}$ Department of Physics, Pusan National University, Pusan, Republic of Korea\\
$^{17}$ Department of Physics, Sejong University, Seoul, Republic of Korea\\
$^{18}$ Department of Physics, University of California, Berkeley, California, United States\\
$^{19}$ Department of Physics, University of Oslo, Oslo, Norway\\
$^{20}$ Department of Physics and Technology, University of Bergen, Bergen, Norway\\
$^{21}$ Dipartimento di Fisica, Universit\`{a} di Pavia, Pavia, Italy\\
$^{22}$ Dipartimento di Fisica dell'Universit\`{a} and Sezione INFN, Cagliari, Italy\\
$^{23}$ Dipartimento di Fisica dell'Universit\`{a} and Sezione INFN, Trieste, Italy\\
$^{24}$ Dipartimento di Fisica dell'Universit\`{a} and Sezione INFN, Turin, Italy\\
$^{25}$ Dipartimento di Fisica e Astronomia dell'Universit\`{a} and Sezione INFN, Bologna, Italy\\
$^{26}$ Dipartimento di Fisica e Astronomia dell'Universit\`{a} and Sezione INFN, Catania, Italy\\
$^{27}$ Dipartimento di Fisica e Astronomia dell'Universit\`{a} and Sezione INFN, Padova, Italy\\
$^{28}$ Dipartimento di Fisica `E.R.~Caianiello' dell'Universit\`{a} and Gruppo Collegato INFN, Salerno, Italy\\
$^{29}$ Dipartimento DISAT del Politecnico and Sezione INFN, Turin, Italy\\
$^{30}$ Dipartimento di Scienze MIFT, Universit\`{a} di Messina, Messina, Italy\\
$^{31}$ Dipartimento Interateneo di Fisica `M.~Merlin' and Sezione INFN, Bari, Italy\\
$^{32}$ European Organization for Nuclear Research (CERN), Geneva, Switzerland\\
$^{33}$ Faculty of Electrical Engineering, Mechanical Engineering and Naval Architecture, University of Split, Split, Croatia\\
$^{34}$ Faculty of Nuclear Sciences and Physical Engineering, Czech Technical University in Prague, Prague, Czech Republic\\
$^{35}$ Faculty of Physics, Sofia University, Sofia, Bulgaria\\
$^{36}$ Faculty of Science, P.J.~\v{S}af\'{a}rik University, Ko\v{s}ice, Slovak Republic\\
$^{37}$ Faculty of Technology, Environmental and Social Sciences, Bergen, Norway\\
$^{38}$ Frankfurt Institute for Advanced Studies, Johann Wolfgang Goethe-Universit\"{a}t Frankfurt, Frankfurt, Germany\\
$^{39}$ Fudan University, Shanghai, China\\
$^{40}$ Gangneung-Wonju National University, Gangneung, Republic of Korea\\
$^{41}$ Gauhati University, Department of Physics, Guwahati, India\\
$^{42}$ Helmholtz-Institut f\"{u}r Strahlen- und Kernphysik, Rheinische Friedrich-Wilhelms-Universit\"{a}t Bonn, Bonn, Germany\\
$^{43}$ Helsinki Institute of Physics (HIP), Helsinki, Finland\\
$^{44}$ High Energy Physics Group,  Universidad Aut\'{o}noma de Puebla, Puebla, Mexico\\
$^{45}$ Horia Hulubei National Institute of Physics and Nuclear Engineering, Bucharest, Romania\\
$^{46}$ HUN-REN Wigner Research Centre for Physics, Budapest, Hungary\\
$^{47}$ Indian Institute of Technology Bombay (IIT), Mumbai, India\\
$^{48}$ Indian Institute of Technology Indore, Indore, India\\
$^{49}$ INFN, Laboratori Nazionali di Frascati, Frascati, Italy\\
$^{50}$ INFN, Sezione di Bari, Bari, Italy\\
$^{51}$ INFN, Sezione di Bologna, Bologna, Italy\\
$^{52}$ INFN, Sezione di Cagliari, Cagliari, Italy\\
$^{53}$ INFN, Sezione di Catania, Catania, Italy\\
$^{54}$ INFN, Sezione di Padova, Padova, Italy\\
$^{55}$ INFN, Sezione di Pavia, Pavia, Italy\\
$^{56}$ INFN, Sezione di Torino, Turin, Italy\\
$^{57}$ INFN, Sezione di Trieste, Trieste, Italy\\
$^{58}$ Inha University, Incheon, Republic of Korea\\
$^{59}$ Institute for Gravitational and Subatomic Physics (GRASP), Utrecht University/Nikhef, Utrecht, Netherlands\\
$^{60}$ Institute of Experimental Physics, Slovak Academy of Sciences, Ko\v{s}ice, Slovak Republic\\
$^{61}$ Institute of Physics, Homi Bhabha National Institute, Bhubaneswar, India\\
$^{62}$ Institute of Physics of the Czech Academy of Sciences, Prague, Czech Republic\\
$^{63}$ Institute of Space Science (ISS), Bucharest, Romania\\
$^{64}$ Institut f\"{u}r Kernphysik, Johann Wolfgang Goethe-Universit\"{a}t Frankfurt, Frankfurt, Germany\\
$^{65}$ Instituto de Ciencias Nucleares, Universidad Nacional Aut\'{o}noma de M\'{e}xico, Mexico City, Mexico\\
$^{66}$ Instituto de F\'{i}sica, Universidade Federal do Rio Grande do Sul (UFRGS), Porto Alegre, Brazil\\
$^{67}$ Instituto de F\'{\i}sica, Universidad Nacional Aut\'{o}noma de M\'{e}xico, Mexico City, Mexico\\
$^{68}$ iThemba LABS, National Research Foundation, Somerset West, South Africa\\
$^{69}$ Jeonbuk National University, Jeonju, Republic of Korea\\
$^{70}$ Johann-Wolfgang-Goethe Universit\"{a}t Frankfurt Institut f\"{u}r Informatik, Fachbereich Informatik und Mathematik, Frankfurt, Germany\\
$^{71}$ Korea Institute of Science and Technology Information, Daejeon, Republic of Korea\\
$^{72}$ Laboratoire de Physique Subatomique et de Cosmologie, Universit\'{e} Grenoble-Alpes, CNRS-IN2P3, Grenoble, France\\
$^{73}$ Lawrence Berkeley National Laboratory, Berkeley, California, United States\\
$^{74}$ Lund University Department of Physics, Division of Particle Physics, Lund, Sweden\\
$^{75}$ Nagasaki Institute of Applied Science, Nagasaki, Japan\\
$^{76}$ Nara Women{'}s University (NWU), Nara, Japan\\
$^{77}$ National and Kapodistrian University of Athens, School of Science, Department of Physics , Athens, Greece\\
$^{78}$ National Centre for Nuclear Research, Warsaw, Poland\\
$^{79}$ National Institute of Science Education and Research, Homi Bhabha National Institute, Jatni, India\\
$^{80}$ National Nuclear Research Center, Baku, Azerbaijan\\
$^{81}$ National Research and Innovation Agency - BRIN, Jakarta, Indonesia\\
$^{82}$ Niels Bohr Institute, University of Copenhagen, Copenhagen, Denmark\\
$^{83}$ Nikhef, National institute for subatomic physics, Amsterdam, Netherlands\\
$^{84}$ Nuclear Physics Group, STFC Daresbury Laboratory, Daresbury, United Kingdom\\
$^{85}$ Nuclear Physics Institute of the Czech Academy of Sciences, Husinec-\v{R}e\v{z}, Czech Republic\\
$^{86}$ Oak Ridge National Laboratory, Oak Ridge, Tennessee, United States\\
$^{87}$ Ohio State University, Columbus, Ohio, United States\\
$^{88}$ Physics department, Faculty of science, University of Zagreb, Zagreb, Croatia\\
$^{89}$ Physics Department, Panjab University, Chandigarh, India\\
$^{90}$ Physics Department, University of Jammu, Jammu, India\\
$^{91}$ Physics Program and International Institute for Sustainability with Knotted Chiral Meta Matter (WPI-SKCM$^{2}$), Hiroshima University, Hiroshima, Japan\\
$^{92}$ Physikalisches Institut, Eberhard-Karls-Universit\"{a}t T\"{u}bingen, T\"{u}bingen, Germany\\
$^{93}$ Physikalisches Institut, Ruprecht-Karls-Universit\"{a}t Heidelberg, Heidelberg, Germany\\
$^{94}$ Physik Department, Technische Universit\"{a}t M\"{u}nchen, Munich, Germany\\
$^{95}$ Politecnico di Bari and Sezione INFN, Bari, Italy\\
$^{96}$ Research Division and ExtreMe Matter Institute EMMI, GSI Helmholtzzentrum f\"ur Schwerionenforschung GmbH, Darmstadt, Germany\\
$^{97}$ Saga University, Saga, Japan\\
$^{98}$ Saha Institute of Nuclear Physics, Homi Bhabha National Institute, Kolkata, India\\
$^{99}$ School of Physics and Astronomy, University of Birmingham, Birmingham, United Kingdom\\
$^{100}$ Secci\'{o}n F\'{\i}sica, Departamento de Ciencias, Pontificia Universidad Cat\'{o}lica del Per\'{u}, Lima, Peru\\
$^{101}$ Stefan Meyer Institut f\"{u}r Subatomare Physik (SMI), Vienna, Austria\\
$^{102}$ SUBATECH, IMT Atlantique, Nantes Universit\'{e}, CNRS-IN2P3, Nantes, France\\
$^{103}$ Sungkyunkwan University, Suwon City, Republic of Korea\\
$^{104}$ Suranaree University of Technology, Nakhon Ratchasima, Thailand\\
$^{105}$ Technical University of Ko\v{s}ice, Ko\v{s}ice, Slovak Republic\\
$^{106}$ The Henryk Niewodniczanski Institute of Nuclear Physics, Polish Academy of Sciences, Cracow, Poland\\
$^{107}$ The University of Texas at Austin, Austin, Texas, United States\\
$^{108}$ Universidad Aut\'{o}noma de Sinaloa, Culiac\'{a}n, Mexico\\
$^{109}$ Universidade de S\~{a}o Paulo (USP), S\~{a}o Paulo, Brazil\\
$^{110}$ Universidade Estadual de Campinas (UNICAMP), Campinas, Brazil\\
$^{111}$ Universidade Federal do ABC, Santo Andre, Brazil\\
$^{112}$ Universitatea Nationala de Stiinta si Tehnologie Politehnica Bucuresti, Bucharest, Romania\\
$^{113}$ University of Derby, Derby, United Kingdom\\
$^{114}$ University of Houston, Houston, Texas, United States\\
$^{115}$ University of Jyv\"{a}skyl\"{a}, Jyv\"{a}skyl\"{a}, Finland\\
$^{116}$ University of Kansas, Lawrence, Kansas, United States\\
$^{117}$ University of Liverpool, Liverpool, United Kingdom\\
$^{118}$ University of Science and Technology of China, Hefei, China\\
$^{119}$ University of South-Eastern Norway, Kongsberg, Norway\\
$^{120}$ University of Tennessee, Knoxville, Tennessee, United States\\
$^{121}$ University of the Witwatersrand, Johannesburg, South Africa\\
$^{122}$ University of Tokyo, Tokyo, Japan\\
$^{123}$ University of Tsukuba, Tsukuba, Japan\\
$^{124}$ Universit\"{a}t M\"{u}nster, Institut f\"{u}r Kernphysik, M\"{u}nster, Germany\\
$^{125}$ Universit\'{e} Clermont Auvergne, CNRS/IN2P3, LPC, Clermont-Ferrand, France\\
$^{126}$ Universit\'{e} de Lyon, CNRS/IN2P3, Institut de Physique des 2 Infinis de Lyon, Lyon, France\\
$^{127}$ Universit\'{e} de Strasbourg, CNRS, IPHC UMR 7178, F-67000 Strasbourg, France, Strasbourg, France\\
$^{128}$ Universit\'{e} Paris-Saclay, Centre d'Etudes de Saclay (CEA), IRFU, D\'{e}partment de Physique Nucl\'{e}aire (DPhN), Saclay, France\\
$^{129}$ Universit\'{e}  Paris-Saclay, CNRS/IN2P3, IJCLab, Orsay, France\\
$^{130}$ Universit\`{a} degli Studi di Foggia, Foggia, Italy\\
$^{131}$ Universit\`{a} del Piemonte Orientale, Vercelli, Italy\\
$^{132}$ Universit\`{a} di Brescia, Brescia, Italy\\
$^{133}$ Variable Energy Cyclotron Centre, Homi Bhabha National Institute, Kolkata, India\\
$^{134}$ Warsaw University of Technology, Warsaw, Poland\\
$^{135}$ Wayne State University, Detroit, Michigan, United States\\
$^{136}$ Yale University, New Haven, Connecticut, United States\\
$^{137}$ Yildiz Technical University, Istanbul, Turkey\\
$^{138}$ Yonsei University, Seoul, Republic of Korea\\
$^{139}$ Affiliated with an institute formerly covered by a cooperation agreement with CERN\\
$^{140}$ Affiliated with an international laboratory covered by a cooperation agreement with CERN.\\

\end{flushleft} 

\end{document}